\documentclass[twocolumn,twocolappendix]{aastex702}

\usepackage{newtxtext,newtxmath}
\usepackage{amsmath}
\usepackage{xspace}

\newcommand{\empdf}{\textsc{emPDF}\xspace}
\newcommand{\opdf}{\textsc{oPDF}}
\newcommand{\Mtwoc}{\ensuremath{M_\text{200c}}\xspace}

\newcommand{\unit}[1]{\ensuremath{\mathrm{\,#1}}\xspace}
\newcommand{\Msun}{\unit{M_\odot}}
\newcommand{\kpc}{\unit{kpc}}

\newcommand{\JointMencEighty}{\ensuremath{6.36^{+0.53}_{-0.48} \times 10^{11} \Msun}}
\newcommand{\JointMtwocRRL}{\ensuremath{1.20^{+0.15}_{-0.13} \times 10^{12} \Msun}}
\newcommand{\JointMtwocBHB}{\ensuremath{1.20^{+0.16}_{-0.14} \times 10^{12} \Msun}}
\newcommand{\JointMtwoc}{\ensuremath{1.19^{+0.14}_{-0.12}\times 10^{12} \Msun}}
\newcommand{\JointMtwocNoGME}{\ensuremath{1.52^{+0.35}_{-0.30}\times 10^{12} \Msun}}

\shorttitle{DESI DR2 Results III: The Mass of the Milky Way}
\shortauthors{DESI Collaboration et al.}

\graphicspath{{./figures/}}

\begin{document}
\title{DESI DR2 Results III: The Mass of the Milky Way}


\author{DESI Collaboration}
\affiliation{Dark Energy Spectroscopic Instrument Collaboration}
\email[show]{spokespersons@desi.lbl.gov}

\author[gname='Christian', sname='Aganze']{C.~Aganze}
\affiliation{Physics Department, Stanford University, Stanford, CA 93405, USA}
\email{caganze@stanford.edu}

\author[gname='Jessica Nicole', sname='Aguilar']{J.~Aguilar}
\affiliation{Lawrence Berkeley National Laboratory, 1 Cyclotron Road, Berkeley, CA 94720, USA}
\email{jaguilar@lbl.gov}

\author[0000-0001-6098-7247, gname='Steven', sname='Ahlen']{S.~Ahlen}
\affiliation{Department of Physics, Boston University, 590 Commonwealth Avenue, Boston, MA 02215 USA}
\email{ahlen@bu.edu}

\author[0000-0002-0084-572X, gname='Carlos ', sname='Allende Prieto']{C.~Allende~Prieto}
\affiliation{Departamento de Astrof\'{\i}sica, Universidad de La Laguna (ULL), E-38206, La Laguna, Tenerife, Spain}
\affiliation{Instituto de Astrof\'{\i}sica de Canarias, C/ V\'{\i}a L\'{a}ctea, s/n, E-38205 La Laguna, Tenerife, Spain}
\email{carlos.allende.prieto@iac.es}

\author[0000-0001-7600-5148, gname='Eric', sname='Armengaud']{E.~Armengaud}
\affiliation{IRFU, CEA, Universit\'{e} Paris-Saclay, F-91191 Gif-sur-Yvette, France}
\email{eric.armengaud@cea.fr}

\author[0000-0001-5998-3986, gname='Alejandro', sname='Aviles']{A.~Aviles}
\affiliation{Instituto Avanzado de Cosmolog\'{\i}a A.~C., San Marcos 11 - Atenas 202. Magdalena Contreras. Ciudad de M\'{e}xico C.~P.~10720, M\'{e}xico}
\affiliation{Instituto de Ciencias F\'{\i}sicas, Universidad Nacional Aut\'onoma de M\'exico, Av. Universidad s/n, Cuernavaca, Morelos, C.~P.~62210, M\'exico}
\email{aviles@icf.unam.mx}

\author[0000-0003-4162-6619, gname='Stephen', sname='Bailey']{S.~Bailey}
\affiliation{Lawrence Berkeley National Laboratory, 1 Cyclotron Road, Berkeley, CA 94720, USA}
\email{stephenbailey@lbl.gov}

\author[0000-0002-9964-1005, gname='Abigail', sname='Bault']{A.~Bault}
\affiliation{Lawrence Berkeley National Laboratory, 1 Cyclotron Road, Berkeley, CA 94720, USA}
\email{abault@lbl.gov}

\author[0009-0002-2434-5903, gname='Jayashree', sname='Behera']{J.~Behera}
\affiliation{Space Telescope Science Institute, 3700 San Martin Drive, Baltimore, MD 21218, USA}
\email{jayashreeb@stsci.edu}

% ...


\begin{abstract}
    We present measurements of the mass of the Milky Way using distant halo stars selected from the Dark Energy Spectroscopic Instrument (DESI) Data Release 2 (DR2).
    We apply four dynamical modeling methods to 10,036 RR Lyrae (RRL) and blue horizontal branch (BHB) stars with 6D phase space coordinates that span from $\sim$10 to 95~kpc from the Galactic center, the largest samples of their kind.
    All methods are applied to mock surveys based on cosmological simulations that include generating individual tracer stars, convolving their phase-space coordinates with DESI and \textit{Gaia} DR3-like uncertainties, and tailored models of the survey selection functions.
    By comparing the performance on these mocks, we quantify how systematic biases from both methodology assumptions and galaxy disequilibrium impact each estimate.
    The mock-based systematics, which differ by method and tracer, are used to correct the individual estimates and to derive joint estimates using a rigorous formalism that accounts for per-method biases and residual correlations between methods.
    We obtain a joint estimate of the mass within 80~kpc to be $M(r < 80\kpc) = \JointMencEighty$ and a virial mass estimate of $\Mtwoc = \JointMtwoc$, with similar results when combining only RRL-based or BHB-based measurements.
    These results provide a roadmap for future DESI efforts, as well as those in other spectroscopic surveys, to reliably constrain the mass of our Galaxy.
\end{abstract}


\section{Introduction} \label{sec:intro}

\setcounter{footnote}{0}

Characterizing the mass distribution of the Milky Way is as scientifically relevant today as it was a century ago when the landmark stellar catalogs of \citet{Kapteyn1922} were published.
While today we have a detailed picture of the distribution of the luminous component \citep{Bland-Hawthorn:2016}\footnote{Though there remain interesting discrepancies in recent measurements of the Milky Way's stellar disk mass \citep{Lian:2025, Imig:2025}.}, we now face the more challenging task of characterizing the total mass, which includes the dark matter distribution within the Galaxy.

After four decades of increasingly compelling evidence, the cold dark matter (CDM) model with a cosmological constant ($\Lambda$) has become the standard framework to describe the evolution of the large-scale structure of the Universe \citep{Davis1985,Weinberg_2013,Planck_2020}.
More recently, apparent shortcomings of the CDM model on galactic and subgalactic scales have been highlighted: the ``missing satellites", ``too-big-to-fail", ``core-cusp" and ``plane-of-satellites" problems \citep[see reviews by][]{Bullock:2017, Sales:2022}.
These small-scale challenges, along with the absence of a detection of the dark matter particle (\citealp{Billard2022}, though see \citealp{Akerib:2026}), have motivated the development of diverse alternative hypotheses for the nature of dark matter that could alleviate the small-scale challenges while retaining the large scale success of CDM \citep[see review by][]{Nadler:2026}.
Examples include, but are not limited to, warm \citep[e.g.,][]{Bode2001, Lovell2014}, self-interacting \citep[e.g.,][]{Spergel2000, Tulin2018}, or fuzzy \citep[e.g.,][]{Hu2000, Schive2014, Hui_2021} dark matter.

Simulations of these alternative models make distinctive predictions for the properties of structure on subgalactic scales, such as the number of dark matter subhalos and satellite galaxies \citep{Springel:2008, Wang2012, Bose:2016, Lovell2016, Enzi2021, Newton2025}, the inner density slope of halos \citep{Maccio2012, Lovell2014, Robertson2019}, the shape of the dark matter halo \citep[e.g.,][]{Dubinski:1991, Bullock:2002, Bose:2016, Robertson2019, Vargya2022} and the presence or absence of figure rotation \citep[][]{Dubinski1994, Bailin2004, Bryan2007, Ash_Valluri_2023}.

A precise determination of the Milky Way's total mass and its spatial distribution are fundamental inputs for constraining dark matter using local data.
For example, the connection between dark matter halos and the galaxies that form within them, and the resulting scaling relations, are determined, to first order, by the halo mass \citep[e.g.,][]{White1991, Cole2000, Behroozi2013, Schaye:2015, Wechsler2018, Schaye:2026}.
For a given halo mass, different model assumptions lead to different predictions for the structure and properties of Milky Way-like galaxies \citep[e.g.,][]{Cooper:2010, Boylan-Kolchin2010, Genina:2023} and the number of subhalos and satellites in a given dark matter model \citep{Kim:2018, Newton:2018, Nadler:2020, Nadler:2021, Nadler:2026}.
Likewise, precisely measuring the local dark matter density is essential for a proper interpretation of direct dark matter detection experiments \citep[see][for a review]{Bertone2005}.
Another example is the reconstruction of the Milky Way's history and the evolution of its different stellar populations, which often requires integrating orbits or calculating integrals of motion in an assumed potential.

Despite its importance, the mass of the Milky Way is still relatively uncertain, with estimates ranging between $\approx 0.5-2\times 10^{12}\Msun$ \citep[see][for reviews]{Bland-Hawthorn:2016, Wang+2020, Bobylev2023}.
In the last decade, data from the \textit{Gaia} mission \citep{Gaia:2016} has stimulated the development, improvement and application of methods to estimate the Milky Way mass distribution \citep[see][for a review]{Hunt+2025}.
A non-exhaustive list of commonly-used methods to estimate the total Milky Way mass includes:
\begin{itemize}
    \item variants of the virial theorem \citep{Evans2003, Watkins2010, Watkins:2019};
    \item application of the Jeans equations in spherical \citep{Gnedin:2010, Kafle+18, Wang+18, Bird:2022} and axisymmetric geometries \citep{Loebman_2014, Wegg_2019, Nitschai_2021};
    \item modeling of stellar streams \citep{Helmi2004, Johnston2005, Law2010, Koposov:2010, Kupper2015, Bovy2016, Bonaca:2018, Erkal:2019, Malhan:2019, Koposov2023, Ibata:2024, Palau2025, Nibauer2025, Bayer:2025};
    \item constraints from the escape velocity \citep{Leonard1990, Smith2007, Piffl2014, Deason:2019, Prudil2022};
    \item distribution function (DF) fitting \citep{McMillan2013, Wang:2015, McMillan:2017,Eadie:2019, Deason2021, Hattori2021, Medina2025c}, including for DFs inferred from simulations \citep{Li2017, Li2019, Li2020, Callingham:2019, Cautun2020}, estimated in the space of integrals in trial potentials \citep{Han_opdf1_2016, LiZhaozhou2025, BeS2025}, or explored via the steady-state collisionless Boltzmann equation \citep{An2021, Green2023}.
\end{itemize}
Finally, there are methods that target the local dark matter density \citep[e.g.,][]{Read:2014,Widmark2021, Buckley2023, Horta2024, Lim2025}, rather than the global mass distribution.

Despite a plethora of methods (and often several implementations of the same method), there is relatively little comparison between methods to assess their relative performance.
Most analyses perform internal validation tests on mock data, but analyses vary on three primary axes: (i) how many modeling techniques are considered, (ii) whether the mock data are based on analytic equilibrium models or cosmological simulations, and (iii) the inclusion of survey effects like selection functions and observational uncertainties.
Several works have applied a single technique to mock data based on cosmological simulations, either with \citep{Li2017, Eadie:2018, Li2019, Zaritsky:2020, Koppelman:2021, Rehemtulla2022, Shen2022_h3mwmass, Medina2025c, Ou:2025} or without \citep{Wang:2015, Sanderson:2017, Callingham:2019, Deason:2019, Grand:2019mwmass_tail, Reino:2022, Oeur:2025, Brooks:2026} survey effects.
Some works have applied multiple techniques to the same dataset, but either used mock data based on equilibrium models \citep{LiZhaozhou2025} or cosmological simulations without survey effects \citep{Han_opdf2_2016, Wang:2017halomodeling, Wang+18}.
Analyses that use nominally similar mocks (e.g., \textit{Gaia} mocks produced in \citealp{Grand:2018} or \citealp{Sanderson:2020}, with additional selections tailored to their data) rarely discuss performance relative to other efforts.

The DESI Milky Way Survey \citep{Cooper_2023} offers an unprecedented opportunity to investigate the Milky Way mass distribution.
We present measurements using distant halo RR Lyrae (RRL) and blue horizontal branch (BHB) stars, with precise 6-D coordinates measured using DESI velocities and \textit{Gaia} proper motions.
Four independent dynamical modeling methods (all assuming spherical symmetry) were applied to $\sim$5200 BHB stars and $\sim$4800 RRL stars with Galactocentric radii from $\sim$10~kpc to 100~kpc, each the largest dataset of its kind.
Each modeling method was validated using the same set of mock surveys generated from cosmological simulations with observational uncertainties and selection effects tailored to the catalogues used in this work.
We then combine these estimates into joint measurements that leverage the relative strengths of each technique across various Galactocentric distances.
This work is accompanied by a set of four supporting papers, described in Table~\ref{tab:supporting-papers}.
The supporting papers describe in detail the mock data sets, the four methods, and the individual mass measurements obtained with each of them.

\begin{deluxetable}{lll}
\tablecaption{Supporting papers relevant to this work and the corresponding sections where their results are discussed.\label{tab:supporting-papers}}
\tablehead{
\colhead{Reference} &
\colhead{Topic} &
\colhead{Sections}
}
\startdata
\citet{Riley:2026auridesi}  & AuriDESI mock catalogues & \ref{sec:mocks} \\
\citet{Zhou:2026empdf}      & \empdf mass constraint    & \ref{sec:empdf-description} \\
\citet{Medina:2026gme}      & GME mass constraint        & \ref{sec:gme-description} \\
\citet{Valluri:2026nimble}  & NIMBLE mass constraints    & \ref{sec:nimble-description} \\
\enddata
\end{deluxetable}

The structure of this paper is as follows: in Section~\ref{sec:data} we describe the DESI DR2 BHB and RRL catalogues, survey selection functions, and mock catalogues used in this work.
Section~\ref{sec:Methods} summarizes the four individual modeling techniques, while Section~\ref{sec:combination} describes our methodology for combining their results into a joint mass estimate.
Section~\ref{sec:mock-results} presents the results of applying the individual modeling techniques and the joint analysis to the mocks.
In Section~\ref{sec:dr2-results} we present the mass measurements for the individual methods and our joint mass determinations using the DESI Milky Way data.
In Section~\ref{sec:discussion} we place our mass measurements in the context of prior literature and explore the implications of our measurements for Milky Way science.
Finally, Section~\ref{sec:summary} presents a summary and our main conclusions.

\section{DESI Data} \label{sec:data}
\subsection{The DESI survey} \label{sec:desi-description}

The Dark Energy Spectroscopic Instrument (DESI) is an extremely efficient, highly multiplexed, fiber-fed spectroscopic facility mounted on the Nicholas U.\ Mayall 4-m telescope at Kitt Peak National Observatory \citep{DESI2016b.Instr, DESI2022.KP1.Instr}.
DESI deploys 5020 robotic fibers over a $3.2^\circ$ diameter field of view at prime focus (i.e., a fiber density of $\approx700~{\rm deg}^{-2}$ \citep{Corrector.Miller.2023, FiberSystem.Poppett.2024}.
The fibers feed ten identical three-arm spectrographs, each of which covers the wavelength range $\lambda\lambda3600-9800$\,\AA\ at high throughput with a spectral resolution varying smoothly from $R \sim 2000$ at the blue end to $5000$ at the red end \citep{DESI2016b.Instr, DESI2022.KP1.Instr, Silber_2023}.
Each fiber can be positioned independently within a patrol region of radius 12~mm ($\approx 1.5$\arcsec), and the entire focal plane can be reconfigured in $\lesssim 2$ minutes, thus enabling rapid tiling of the sky and efficient spectroscopic surveys \citep{SurveyOps.Schlafly.2023, FiberSystem.Poppett.2024}.
The data processing pipeline reduces and analyzes each night's data by the following morning \citep{Spectro.Pipeline.Guy.2023}.
DESI's primary 8-year survey (2020-2028) will map 17,000~deg$^2$ and measure accurate spectroscopic redshifts of nearly 63 million galaxies and quasars and radial velocities for more than 20 million stars.

DESI's primary scientific mission \citep{DESI_2016a} is to place percent-level constraints on the parameters of the equation of state of dark energy using the baryon acoustic oscillation method (which measures the redshift variation of the angular diameter distance and the Hubble constant out to redshift $z\gtrsim 3.5$) and redshift space distortions (which measure the growth of structure and probe potential modifications to general relativity).
In addition, DESI will constrain the sum of the neutrino masses, study the scale dependence of primordial density fluctuations from inflation, and test potential modifications to the general theory of relativity.
Early results already reveal tantalizing hints that dark energy may be dynamic and that the standard model may be different from $\Lambda$CDM \citep{DESI_DR1_Cosmology, DESI2024.VII.KP7B, DESI.DR2.DR2}.

In parallel with its cosmological goals, DESI is also conducting a survey of stars in the Milky Way and Local Group.
During bright observing conditions when high-redshift observations are inefficient, the DESI Milky Way Survey \citep[MWS;][]{Cooper_2023} has obtained spectra for $\approx 10$ million stars across $\sim 14,000$~deg$^2$.
In addition, the DESI Milky Way Backup Program \citep[MWBP;][]{DESI_Backup_Program} has measured spectra of an additional $\approx$10 million stars over a wider footprint of nearly 21,000~deg$^2$.
In addition to these two programs, DESI is also undertaking more focused programs targeting a large number of Milky Way streams \citep[e.g.,][]{Mohammed:2026, Jarvis:2026}, globular clusters and dwarf satellite galaxy companions \citep[e.g.,][]{Ding:2025, Yang:2025}, and a survey of the Andromeda Galaxy \citep{DESI_Andromeda_2023}.
Together, these constitute one of the largest spectroscopic surveys of faint stars in the Milky Way and its closest neighbors. 

The DESI MWS is designed to obtain a large, minimally biased spectroscopic sample of stars primarily at Galactic latitudes $|b| > 20^\circ$, targeting the thick disk and stellar halo \citep{Cooper_2023}.
The MWS operates during bright time and shares the focal plane with the DESI Bright Galaxy Survey \citep[BGS;][]{DESI_BGS_2023}, utilizing otherwise unassigned fibers to efficiently observe stellar targets \citep{SurveyOps.Schlafly.2023}.
The primary MWS sample is defined over the magnitude range $16 < r < 19$ and is divided into three complementary target classes --- \texttt{MAIN-BLUE}, \texttt{MAIN-RED}, and \texttt{MAIN-BROAD} --- that together provide near-complete coverage of color--magnitude space \citep[see][for more details]{Cooper_2023}.
These classes are designed to ensure an inclusive selection function that is amenable to forward modeling, while maintaining sensitivity to different stellar populations:
\begin{itemize}
    \item \texttt{MAIN-BLUE} selects blue stars ($g-r < 0.7$), dominated by metal-poor turnoff stars and horizontal branch stars.
    \item \texttt{MAIN-RED} uses \textit{Gaia} astrometry to preferentially select distant halo giants among redder stars.
    \item \texttt{MAIN-BROAD} includes remaining red stars at lower priority, primarily sampling disk populations.
\end{itemize}
In addition to these three target classes, MWS also includes several high-priority subsamples of rare stellar tracers, including BHBs and RRLs that are the key focus of this paper.

The DESI nightly observations are reduced by an automated spectroscopic pipeline \citep{Spectro.Pipeline.Guy.2023} and, periodically, collected into a data set which is jointly re-processed in a uniform manner and released publicly.
The Early Data Release \citep[EDR;][]{DESI_EDR_2024} consists of all the data obtained during the initial Survey Validation period.
Data Release 1 \citep[DR1;][]{DESI2024.I.DR1} replaces the EDR, and contains $\approx$18.7 million spectra, including measurements of more than 4 million stars \citep[see][for the DR1 Stellar Catalog]{Koposov_2026}. 
This paper makes use of private data from Data Release 2 (DR2), which covers the first three years of DESI observations.
DR2 contains more than 30 million extragalactic spectra and more than 11.8 million stellar spectra (6.25 million from the main MWS, 4.85 million from the MWBP, and 0.7 million from the dark time program).

\subsection{BHB catalogue} \label{sec:bhb-catalogue}

Blue horizontal branch (BHB) stars are excellent distance tracers of the outer Milky Way halo. 
They are luminous and can thus be observed at large Galactocentric distances, while their absolute magnitude can be computed using only their observed colour to yield reliable distances.
To measure the mass of the Milky Way in this work, we use an updated version of the BHB catalogue originally described in \citet{Bystroem2025_BHBcatLMCwork}.
The steps taken to create the catalogue are described in that paper, but we summarise the key steps here.

BHB sources are given high fiber assignment priority than the main MWS sample, and the targeting is based on the main sample with additional requirements.
More information about this selection can be found in \citet{Cooper_2023}.
The paper by \citet{Bystroem2025_BHBcatLMCwork} used BHBs observed in both bright and dark time, but to simplify the BHB selection function required in our modelling, this work only uses bright time BHB observations.

Since the BHB targeting is based on photometry and astrometry alone, the sample needs to be cleaned from contamination.
To achieve a clean sample of BHB stars we follow the description in \citet{Bystroem2025_BHBcatLMCwork}, with two changes.
First, we require that all objects are flagged as stars by the template-fitting code Redrock (Bailey et al. in prep.) by using the selection \texttt{RR\_SPECTYPE = STAR} instead of only removing quasars.
Second, we require all stars to satisfy \texttt{RVS\_WARN = 0}, which removes all objects with error flags from the RVS pipeline \citep{Koposov2024, Koposov_2026}, ensuring trustworthy stellar parameters and radial velocities.
From the resulting stellar sample we remove RRLs by crossmatching the sample with RRL catalogs from the literature.
We also remove blue stragglers by applying a cut in stellar surface gravity and effective temperature on the remaining stars (see Equation~5 of \citealp{Bystroem2025_BHBcatLMCwork}).
This leaves us with a sample of spectroscopically confirmed BHBs, with no obvious outliers in stellar parameters or kinematics (e.g., phase-space diagram $r$ vs.~$v_r$).

To compute the absolute magnitude of the BHBs, we use the absolute magnitude-colour relation in Equation~6 in \citet{Bystroem2025_BHBcatLMCwork}, which is only valid in the colour range $-0.3 < (g_0-r_0) < 0.0$.
BHBs outside this range are also removed from the sample.
From the resulting distances, we find that the relative distance error for each BHB is 5.7~per~cent.
This is calculated by following the distance validation in \citet{Bystroem2025_BHBcatLMCwork}, where we use the GC members from \citet{vasiliev_GCs} and add dwarf galaxy members from \citet{Pace2022}, and compute the standard deviation of the ratio of our BHB distances and the literature values.

\begin{figure}
\centering
\includegraphics[width=\linewidth]{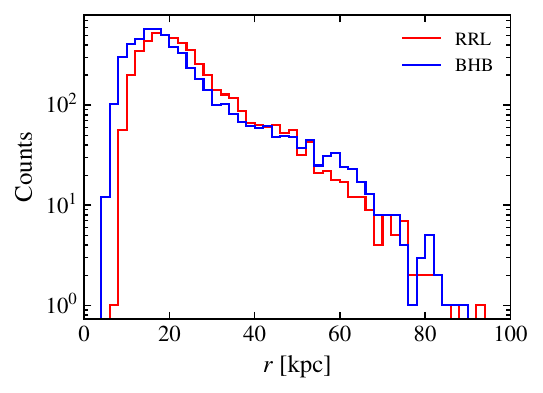}
\caption{Galactocentric radii of the DESI BHB and RRL samples used in this work.
Note that individual methods may apply additional cuts to the data, see Table~\ref{tab:supporting-selections}.}
\label{fig:bbh_rrl_distances}
\end{figure}

The final step is to remove known halo substructures from the sample, as these kinematic substructures are known to bias mass estimates.
Dwarf galaxies and globular clusters are removed by crossmatching the sample with catalogues of such objects \citep{vasiliev_GCs, Pace2022}.
The remaining dominant substructure is the prominent Sagittarius (Sgr) stellar stream \citep{Ibata:2001sgr_carbonstars, Newberg:2002, Majewski:2003}, which we remove by excluding all stars that have stream latitude $|B_\text{Sgr}| < 15\deg$ and have distances within $3\sigma$ of the Sgr distance track of \citet{hernitschek17_Sgrdistancesplines}.
This cut is given in Equation~7 and shown in Figure~3 of \citet{Bystroem2025_BHBcatLMCwork}.
After all cuts are applied, there are 5,198 BHB stars remaining in the fiducial sample.
The Galactocentric distance distribution of these BHBs is shown in Figure~\ref{fig:bbh_rrl_distances}.

\subsection{RRL catalogue} \label{sec:rrl-catalogue}

Similar to BHB stars, RRL stars are considered one of the best tracers of the outer halo, due to their ubiquity in those regions and their status as precise distance indicators \citep[with typical uncertainties $<5$~per~cent; see e.g.,][]{Catelan2015,Medina2024}.
RRL targets in DESI MWS are mainly observed as part of the survey's BRIGHT and DARK programs \citep[see e.g.,][]{Cooper_2023,Koposov2024}. 
In this work, we use the DESI DR2 RRL catalogue, which expands the sample presented by \citet{Medina2025_y1rrls_paper1} using DESI DR1 from 6,240 to 12,868 RRLs.
This updated sample, which results from a crossmatch between observed DESI targets and the \textit{Gaia} RRL catalogue \citep[][]{Clementini2023}, consists of a total of 28,410 individual spectroscopic epochs, and reach distances out to $\sim120$\,kpc from the Galactic centre. 
The number of spectroscopic epochs per star ranges from 1 to 48, with the bulk of the outer halo sample having fewer than 10 observations.
The selection and analysis of RRLs in DESI DR2 follows the methodology adopted by \citet{Medina2025_y1rrls_paper1}, and will be described in detail in a separate paper. 
Here we summarize the most important aspects of their analysis.

Heliocentric distances for RRLs are computed from the absolute magnitude--metallicity relation of \citet{Garofalo2022} in the \textit{Gaia} $G$~band, using the intensity-averaged $G$~magnitude ({\tt INT\_AVERAGE\_G}) from the \textit{Gaia} source catalog and iron abundances derived from DESI spectra and averaged across epochs. 
Extinction corrections are applied using the dust maps of \citet{Schlafly:2011}, adopting $A_G/A_V = 0.85926$.
The typical distance uncertainty is $\sim$5~per cent, driven primarily by the scatter in [Fe/H] \citep[see][]{Medina2025_y1rrls_paper1}. 
The mean signal-to-noise ratio of individual DESI spectra declines with distance and falls below $\sim$10 for most stars at Galactocentric distances beyond $\sim$50~kpc. 

Systemic radial velocities (i.e., center-of-mass velocities) are derived following the Bayesian radial velocity curve (RVC) modeling methodology described in \citet{Medina2025_y1rrls_paper1}. 
For each star, the pulsation phase of every spectroscopic observation is estimated from the \textit{Gaia}~DR3 period and reference epoch, and a Fourier series model is simultaneously fit across all available epochs (combining every star in our sample) to infer the systemic velocity $v_{\rm sys}$ of each star.
The model is implemented in Stan using the Python interface CmdStanPy\footnote{\href{https://github.com/stan-dev/cmdstanpy}{https://github.com/stan-dev/cmdstanpy}} \citep{stan-manual}.
Section~4 of \citet{Medina2025_y1rrls_paper1} validates the line-of-sight velocities against external measurements.
Resulting velocity uncertainties are $<$10~km\,s$^{-1}$ for the majority of our sample ($\sim$70~per~cent).
RRLs with uncertainties above this limit typically have fewer DESI epochs ($N \lesssim 5$), epochs that occur at RRL phases near maximum light ($\phi < 0.1$ or $\phi > 0.9$), or both.

For this work, we construct a clean sample of halo field RRLs excluding stars associated with known globular clusters or dwarf galaxies through a positional crossmatch with the catalogs of \citet{vasiliev_GCs} and \citet{Pace2022}, using a matching radius of 0.5~arcsec \citep{Medina2025_y1rrls_paper1}. 
Additionally, we remove candidate members of the Sgr stream from our sample using the same procedure as described for the BHBs in the previous subsection.

Lastly, we require $5{,}000 < T_{\rm eff}/{\rm K} < 8{,}500$ and $0 < \log g/{\rm dex} < 3.8$ (as reported by the RVS pipeline) for the RRLs in our sample, since stars outside these ranges appear as clear outliers in Kiel space and are likely contaminants. 
Two further stars (\textit{Gaia}~DR3 {\tt source\_id}s 2708106729874816896 and 382100910792267136) are excluded on the same basis (i.e., they are outliers in $T_{\rm eff}$--$\log g$ space).
We also remove stars brighter than $G_0 = 16$, to more closely match the magnitude distribution of the BHB sample.
After all selection cuts, our fiducial RRL sample comprises 4,838 stars, which is the sample used for dynamical modeling in this work. 
The Galactocentric distance distribution of these RRLs used is shown in Figure~\ref{fig:bbh_rrl_distances}.
Similar to the BHB sample, there are no obvious outliers in stellar parameters or Galaxy-wide kinematics.

\subsection{Selection functions} \label{sec:selection-functions}

\begin{figure}
\includegraphics[width=\linewidth]{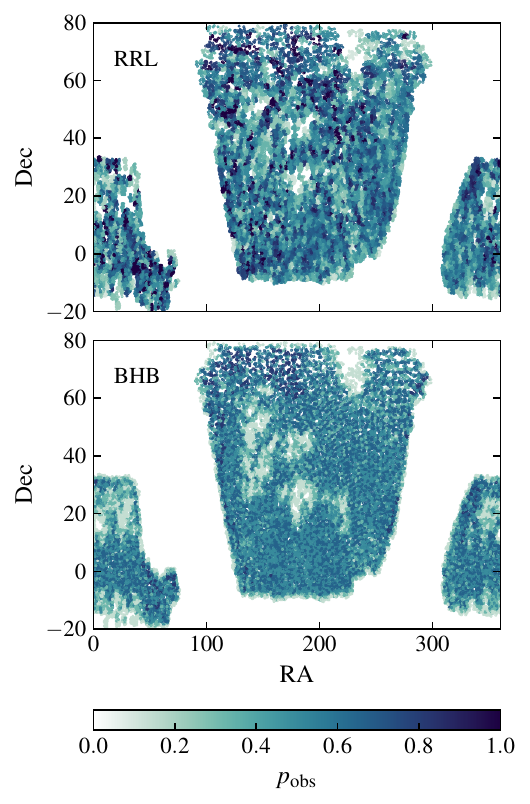}
\caption{Selection functions of the DESI RRL (top) and BHB (bottom) samples as a function of sky position.
The BHB selection function is parameterized by the number of times a HEALPix pixel has been observed by DESI.
In contrast, the RRL selection function is computed directly on the DESI observing tile grid.}
\label{fig:selection_functions}
\end{figure}

DESI is an ongoing spectroscopic campaign with a finite number of fibers and opportunities to (re-)visit a particular location in the sky.
As such, quantifying variation in completeness within the survey footprint after three years of observations is an important ingredient of our mass modeling effort.
Analogous to the DESI MWS K-giant sample \citep{Li:2026}, the selection functions of the DESI MWS BHB and RRL samples show no significant dependence on color or apparent magnitude.
As such, we treat both selection functions as exclusively dependent on sky position.

Figure~\ref{fig:selection_functions} presents the selection functions of the DESI MWS RRL (top panel) and BHB (bottom panel) samples as a function of sky position.
The BHB selection function is computed directly within HEALPix pixels, with a HEALPix resolution parameter\footnote{This resolution partitions the sky into 12288 equal-area pixels, each with an angular area of $5.3 \, \mathrm{deg}^2$.} of $N_{\rm side} = 32$.
Within a set of pixels that have been observed the same number of times by DESI, the completeness of BHB targets is roughly uniform.
We compute the mean completeness for pixels with the same number of visits, then assign that completeness to all such pixels.
We assume that our BHB target catalogue is complete to our magnitude limit, which is reasonable given the quality of the Legacy Survey imaging (limit of $r=23.9$, uncertainties of $\lesssim$0.003 mag at $r=20$, \citealp{Dey:2019}) that the selection is based on.

In contrast, the RRL selection function is computed on the DESI observing tile grid, and is defined simply as the ratio of observed to target stars in a given tile.
Since RRLs are pulsating stars with variable radial velocities, in DESI they are re-observed many times in overlapping, offset layers of exposures.
This simpler selection function model helps empirically account for the fact that RRL completeness within a patch of sky depends on the SNR of individual epochs and that there is a stronger preference to obtain multiple epochs for RRLs than for BHBs.
Additionally, the \textit{Gaia} RRL catalogue used for targeting is not as complete as the BHB target catalogue, due to \textit{Gaia}'s brighter limiting magnitude and scanning law.
To account for this, we use the empirical \textit{Gaia}~DR2 RRL selection function determined by \citet{Mateu:2020}, which depends on \textit{Gaia} $G$-band magnitude in addition to sky location.
The selection function used for the mock catalogues and reported for the real DESI RRLs includes both the DESI and \textit{Gaia} completeness factors, while Figure~\ref{fig:selection_functions} only shows the DESI component.

Both the DESI DR2 MWS BHB and RRL samples achieve high completeness over the observed sky.
The median value of the selection function across all pixels is $\sim$0.5 for both samples, indicating a high and consistent observational efficiency for the two stellar populations.
Both selection functions have large-scale regions of lower completeness within the survey footprint, corresponding to few or no DESI visits.
Within regions of higher completeness, the RRL selection function shows fluctuations on smaller angular scales that are primarily due to the implementation differences described above.

\subsection{Mock catalogues} \label{sec:mocks}

We use mock star catalogs (hereafter `mocks') derived from the Auriga suite of cosmological simulations \citep{Grand:2018, Kizhuprakkat:2024} to validate and interpret the results of our mass modeling techniques.
The underlying simulations follow the growth of the halo and central galaxy self-consistently, with each halo experiencing a unique accretion history.
The mocks therefore capture the non-equilibrium dynamics of the stellar halo and allow us to evaluate systematic biases in our mass estimates due, for example, to the presence of substructure in the phase space distribution of tracer stars \citep[e.g.,][]{Han_opdf1_2016}.
This may be particularly important for the debris of the most massive stellar halo progenitors, which are likely to be responsible for the most prominent phase-space substructures in the Milky Way (Sgr and GSE; \citealp{Helmi:2018, Belokurov:2018}).

The orbital evolution (and corresponding mass loss histories) of massive progenitors are expected to be complex and sensitive to the gravitational response of the central potential, including the disk \citep{Laporte2018, Vasiliev:2022aa}.
The stellar halo may also show spatial and kinematic asymmetries that are signs of on-going interactions with massive satellites like the LMC \citep[e.g.,][]{Gomez:2015, Garavito-Camargo:2019, Cunningham:2020, Petersen:2021, Erkal:2021, Conroy:2021, Yaaqib:2024, Bystroem2025_BHBcatLMCwork, Chandra:2025}.
Such effects cannot be  captured by dynamical models that treat each accretion event in isolation and assume an analytically time-dependent but otherwise rigid central potential \citep[for example,][]{Bullock:2005}.
In principle, \textit{ab initio} simulations also better account for systematics due to the survey selection function, because the phase space distribution of particular stellar populations is likely sensitive to the star formation histories of individual dominant progenitors.

The advantages of mocks based on cosmological simulations come with a significant caveat -- those simulations are subject to a great deal of modeling uncertainty.
Simulations can provide robust realizations of \textit{total} mass assembly histories and satellite dynamics for halos of a given present-day virial mass.
However, the baryonic physics in those simulations, and hence their predictions for galaxy formation, are much less well constrained.
For example, the simulations may produce over-massive stellar disks in the host galaxy, and over-massive progenitor satellites, leading to correspondingly massive and metal-rich stellar halos \citep{Cooper:2015aa, Cooper:2025aa, Deason:2016, Monachesi:2019, Thomas:2021}.
For the work we present here, it is not necessary that the mocks closely resemble the real Milky Way in every respect.
Indeed, to be robust to our lack of knowledge regarding the properties and history of the Milky Way itself, useful methods for estimating the virial mass from stellar dynamics should work equally well for systems with a range of different mass assembly and star formation histories.

The AuriDESI mocks \citep{Kizhuprakkat:2024} provide realizations of a DESI-like survey in six isolated, late-type galaxies from resolution level 3 of the Auriga suite of cosmological magnetohydrodynamical simulations \citep{Grand:2017,Grand:2024} in a \citet{PlanckCollaboration:2014} cosmology.\footnote{$\Omega_\mathrm{m} = 0.307$, $\Omega_{b} = 0.048$, $\Omega_{\Lambda}= 0.693$, $\sigma_8 = 0.8288$, and $H_0 = 100 h \,\mathrm{km\,s^{-1}\, Mpc^{-1}}$ with $h = 0.6777$.}
The Auriga galaxies have virial masses in the range $1 < M_\mathrm{200c} < 2 \times 10^{12}\,\mathrm{M_\odot}$ and total stellar masses in the range $6 < M_\mathrm{\star} < 10 \times 10^{10}\,\mathrm{M_\odot}$.
As with the earlier AuriGaia mocks \citep{Grand:2018}, the AuriDESI mocks uses the \citet{Lowing:2015} phase-space kernel sampling method\footnote{Itself a development of the techniques presented in \citet{Sharma:2011_galaxia} and \citet{SharmaSteinmetz:2011}.} to assign coordinates and velocities to mock stars.
The stellar masses and corresponding atmospheric parameters are sampled from the PARSEC model isochrones with a \citet{Chabrier:2003} IMF.
AuriDESI provides\footnote{The AuriDESI data are available at \url{https://data.desi.lbl.gov/public/papers/mws/auridesi/v1/}} mock realizations of both the DESI MWS target catalogs and the final spectroscopic data products.
For each halo, the mocks provide survey realizations viewed from four different solar azimuthal positions in the simulated galaxy (viewing angles $30^\circ$, $120^\circ$, $210^\circ$ and $300^\circ$, relative to the major axis of the galactic bar).

For our mass-model comparisons, we have constructed BHB- and RRL-specific mocks based on AuriDESI; these are described in \citet{Riley:2026auridesi}.
In particular, these mocks assign measurement uncertainties tailored to the fiducial BHB and RRL samples and incorporate the selection functions presented in Section~\ref{sec:selection-functions}.
Analogous to the fiducial sample construction, we also identify (and remove) stars bound to satellites and from massive stellar streams (analogous to the Sgr stream), but keep stars that originate from phase-mixed tidal debris (e.g., GSE) or cold streams that are less massive than Sgr (e.g., Orphan-Chenab).

Although they are broadly similar, the six AuriDESI systems have somewhat different masses, mass assembly histories and central galaxy properties, as described in more detail by \citet{Grand:2018} and \citet{Kizhuprakkat:2024}; see also \citet{Monachesi:2019}, \citet{Fattahi:2020}, \citet{Orkney:2023}, \citet{Pu:2025aa} and \citet{Thomas:2025} for further discussion of the stellar halos in these galaxies.
Au-6 is the most Milky Way-like in several respects: it has the lowest total mass, $M_\mathrm{200c} \simeq 1.0\times10^{12}\Msun$, although its stellar disk ($M_\mathrm{\star,d} \simeq 2.6\times10^{10}\Msun$) is approximately half the mass of the Milky Way disk, and hence much closer to the abundance-matching mean \citep[e.g.][]{Moster:2013}.
Au-6 accretes only three satellites more massive than $5\times10^{9}\Msun$ in the last 8 Gyr: the most massive ($\Mtwoc\sim10^{11}\,\Msun$) is accreted at $t_\text{lbk}\sim5\,\mathrm{Gyr}$, and has a wide orbit with a pericenter of $\sim$100 \kpc; the other two (both $\Mtwoc \sim10^{10}\,\Msun$) are accreted at $t_\text{lbk}\sim7\,\mathrm{Gyr}$ and have pericenters of 73 and 25~kpc, the latter losing more than 90~per~cent of its initial mass. 

In contrast, in Au-21, a satellite of $\Mtwoc \sim10^{11}\,\Msun$ accreted at $t_\text{lbk}\sim7\,\mathrm{Gyr}$ has a pericenter of $\sim$40~kpc and a present-day galactocentric radius of $\sim$70~kpc.
This progenitor loses 90~per~cent of its total mass.
Its debris boosts the mass of the accreted stellar halo by a factor of $\sim$2-3 relative to the other halos, and the total number of stars in the target catalog of the mock AuriDESI observer by to a factor of $\sim$10 at $\sim$70~kpc (see Figure~3 of \citealt{Kizhuprakkat:2024}; this is Progenitor 4 in Figure~D1 of that paper).
Similar cases occur in Au-23 (where the satellite is fully disrupted before $t_\text{lbk}\sim4\,\mathrm{Gyr}$ and Au-27 (where the satellite survives).
Thus, Au-21, Au-23 and Au-27 may be considered to have the most `disturbed' stellar halos, in terms of the prominence of unrelaxed (stream-like) material.
Figure~1 of \citet{Riley:2025} provides a visual summary of the distribution of material in different states of relaxation in each halo.
However, we caution that it is by no means straightforward to infer from a list of progenitor masses, infall times and pericenters the likely degree of `disturbance' to the present-day stellar halo, and hence the possible impact on dynamical estimates of the total halo mass.

\section{Dynamical Modeling Methods} \label{sec:Methods}

\begin{deluxetable*}{llccl}
\tablecaption{Additional cuts that individual methods apply to the fiducial BHB (Section~\ref{sec:bhb-catalogue}) and RRL (Section~\ref{sec:rrl-catalogue}) samples, the number of BHBs and RRLs actually modeled, and the parameters that each method fits. Selection criteria are discussed in the respective supporting papers.}
\label{tab:supporting-selections}
\tablehead{
\colhead{Method} &
\colhead{Inclusion criteria} &
\colhead{$N_\text{BHB}$} &
\colhead{$N_\text{RRL}$} &
\colhead{Fitted parameters}
}
\startdata
\empdf   & $r > 20 \kpc$ & 1863 & 2282 & Halo mass $M_\text{200c}$, concentration $c_\text{200c}$ \\
& $16 < r_0 < 20$ \\
& $|B_\text{Sgr}| > 15^\circ$ for all distances \\
\hline
GME      & $r > 50 \kpc$ & 352 & 213 & Potential norm $\Phi_0$, potential slope $\gamma$, tracer slope $\alpha$, anisotropy $\beta$ \\
& $G < 20$ \\
\hline
NIMBLE-A & $10 < r < 80 \kpc$ & 4628 & 4798 & B-splines for velocity dispersions and number density \\
& $16 < G < 20$ \\
& $v_\text{tan} < 600$~km~s$^{-1}$ for $r > 50 \kpc$ \\
\hline
NIMBLE-B & $r > 10 \kpc$ & 4573 & 4371 & Two B-splines (halo and outlier) for velocity dispersions and means
\enddata
\end{deluxetable*}

We use four independent dynamical modeling methods, described in detail in the following sections.
We present the results of applying these methods to the mocks in Section~\ref{sec:mock-results}, and the Milky Way mass measurements using the DESI Milky Way data in Section~\ref{sec:dr2-results}.
All methods assume spherical symmetry and dynamical equilibrium, but fit the data very differently.
Table~\ref{tab:supporting-selections} summarises the additional data cuts each method applies (see the supporting papers for details) and the parameters each fits.
All methods return a mass enclosed profile as a function of radius, $M(<r)$.

\subsection{emPDF} \label{sec:empdf-description}

\begin{figure}
\centering
\includegraphics[width=\linewidth]{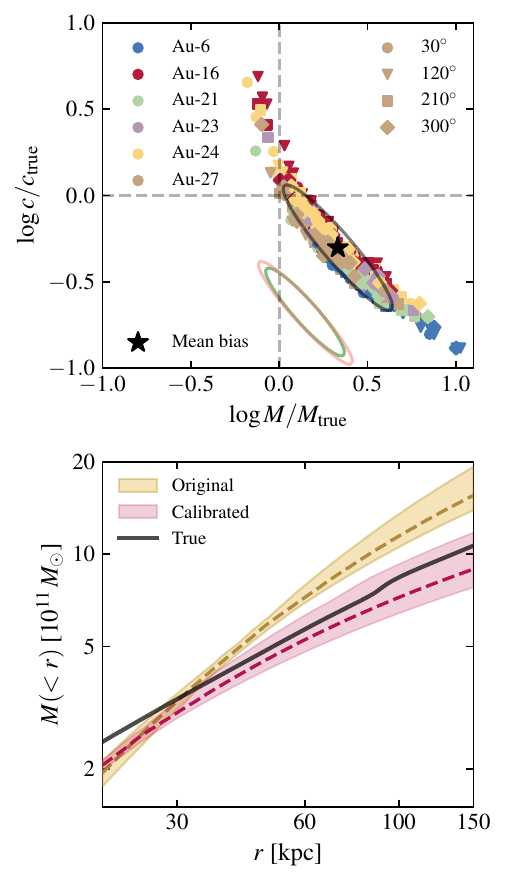}
\caption{Calibration procedure of the \empdf method.
{\bf Top:} Best-fit halo mass and concentration from the mock BHB data versus the true values.
Colors denote the simulated halo, while symbols denote the viewing angle.
The green ellipse represents the average size of $1\sigma$ formal error for each realization.
The pink ellipse represents the size of the $1\sigma$ statistical error for the 20 realizations, averaged over halo-view combinations.
The gray ellipse marks the $1\sigma$ total scatter of all halos, viewings, and realizations.
The green and pink ellipses are placed arbitrarily, while the gray ellipse is centered on the ensemble-averaged bias adopted for the calibration (black star).
{\bf Bottom:} Example enclosed mass profile for the BHB catalogue for Au-21 at the 30$^\circ$ viewing.
The true enclosed mass profile of the halo is shown in black, while the original \empdf output and calibrated results are shown in brown and red.
These calibrated results, which are the \empdf input to the joint estimate, include the statistical uncertainty only (see text for details).
}
\label{fig:empdf_example}
\end{figure}

\empdf (Empirical Distribution Function, \citealp{LiZhaozhou2025}) is an extension of the orbital Probability Density Function (\opdf, \citealp{Han_opdf1_2016,Han_opdf2_2016}) method, which allows for the construction of a distribution function from the data itself, essentially only assuming the tracer distribution to be in a steady-state.

In general, a phase-space distribution function can be fully decomposed into a distribution of orbital parameters, along with the distribution of particles on each orbit.
In the case of a spherical system, it is sufficient to specify the orbits via energy and angular momentum, such that
\begin{equation}
f(\mathbf{r,v})=\frac{|v_r|}{8\pi^2L}p(r|E,L)p(E,L).
\end{equation} 
Here $f(r,v)$ is the phase-space DF, $p(E,L)$ is the distribution of energy and angular momentum, and $p(r|E,L)$ specifies the conditional distribution along each orbit. 
For any assumed gravitational potential, the distribution of orbital parameters $p\left(E,L\right)$ can be readily converted from the observed phase-space coordinates of the tracer particles.
To suppress shot noise in the derived distribution function, \empdf applies a kernel density estimation (KDE) technique to derive a smooth estimation of $p(E,L)$, weighting by the selection function and a smoothing kernel adaptively determined from the data distribution~\citep[see][for details]{LiZhaozhou2025}.

With the DF constructed empirically as above, the probability of observing a star with position and velocity $(\boldsymbol{r}_i,\boldsymbol{v}_i)$ under a model potential, $\Phi(r)$, specified by parameters $\boldsymbol{\Theta}_\Phi$, is the DF itself, $p(\boldsymbol{r}_i,\boldsymbol{v}_i|\boldsymbol{\Theta}_\Phi)=f(\boldsymbol{r}_i,\boldsymbol{v}_i)$.
The posterior of the model potential parameters for the full sample of tracer stars is then
\begin{equation}
    p(\boldsymbol{\Theta}_\Phi|\{\boldsymbol{r}_i,\boldsymbol{v}_i\})\propto p(\boldsymbol{\Theta}_\Phi)\prod_{i=1}^Np(\boldsymbol{r}_i,\boldsymbol{v}_i|\boldsymbol{\Theta}_\Phi),
\end{equation}
where $p(\boldsymbol{\Theta}_\Phi)$ is the prior on the potential parameters.
By maximizing the posterior, the best-fit potential can be obtained.
This model parameterizes the Galactic potential as a spherical NFW profile \citep{Navarro:1996}.

The current version of \empdf does not model observational errors.
In order to properly account for the effect of observational errors, which would artificially increase the velocity dispersion and bias inferred mass to be higher, we include corrections determined using an additional set of AuriDESI mock catalogs \citep{Zhou:2026empdf}.
Briefly, \empdf is applied to six Milky Way-like galaxies from the Auriga suite each viewed at four different viewing angles, for a total of 24 DESI-like mocks.
For each halo and viewing angle, there are 20 observational realizations of AuriDESI that are generated in the same manner as the fiducial set, a total of (24$\times$20) error-convolved mock realizations.

As seen in the top panel of Figure~\ref{fig:empdf_example}, the best-fit halo mass and concentration are biased (as expected).
The {\it ensemble averaged} bias between the best-fit and true halo parameters over all these mocks is adopted for calibration.
More clearly, for each mock and for the real DESI data, the best-fit output from the \empdf method is shifted by the mean bias in the $(\Mtwoc, c_\text{200c})$ plane (black star), with slightly different shifts for BHB and RRL.
The scatter of the fitted parameters among the 20 realizations (pink ellipse in Figure~\ref{fig:empdf_example}) provides an empirical estimate of the statistical uncertainty for a fixed halo and viewing angle; this scatter is comparable to the formal uncertainties output by \empdf (green ellipse).
The total scatter among the inferred parameters across all 480 mock systems (black ellipse) is larger than the formal error.

The results of the calibration can be seen in the bottom panel of Figure~\ref{fig:empdf_example}, an example fit on the mock BHB system of Au-21 with viewing angle of 30$^\circ$ with respect to the bar major axis.
The brown dashed curve with shaded error band shows the enclosed mass profile before applying the bias correction.
The red dashed curve with the shared error band shows the calibrated mass profile and statistical error -- which is shifted toward the true profile (black solid curve).
These calibrated results are the \empdf input to the joint estimate derived in Section~\ref{sec:combination}.

In addition, \citet{Zhou:2026empdf} estimate the systematic covariance component based on the scatter of all mock catalogs after subtracting the formal covariance, and then add it back to the formal covariance of \empdf when applying to observational data and to each individual mock system.
This `total' uncertainty is effectively captured in our joint estimate methodology (but modeled consistently with other methods), so we exclude it in this work.

\subsection{GME} \label{sec:gme-description}

\begin{figure}
\centering
\includegraphics[width=\linewidth]{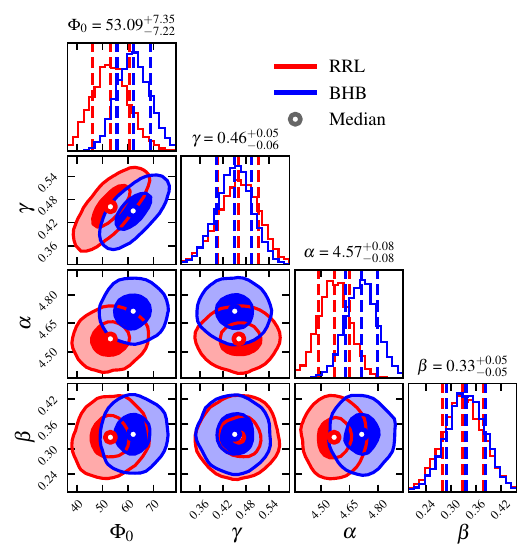}
\caption{Corner plot showing the marginal posterior probabilities and correlations for the GME model parameters applied to the DESI DR2 RRLs (red) and BHBs (blue).
Vertical dashed lines show the 16, 50, and 84 percentiles, which are also shown in text above the 1-D histograms for the RRL results.}
\label{fig:gme_params}
\end{figure}

GME (Galactic Mass Estimator) is a distribution function fitting method that uses a hierarchical Bayesian framework to infer the Milky Way mass from the positions and velocities of halo tracers \citep{Eadie2015, Eadie:2019, Shen2022_h3mwmass, Medina2025c}.
This method assumes that the total potential of the Milky Way, $\Phi(r)$, is modeled as a single power law, with respect to the Galactocentric radius $r$,
\begin{equation}
\Phi(r) = \Phi_0\ r^{-\gamma},
\end{equation}
where $\Phi_0$ is a scale factor, and $\gamma$ is the power-law index.
Tracers are assumed to be in steady state and have radial density profile, $\rho(r)$, also described by a single power law with index $\alpha$,
\begin{equation}
\rho \propto r^{-\alpha}.
\end{equation}
Further with the adoption of a constant velocity anisotropy ($\beta$) across the probed distance range, the probability of any particle being at a given location in phase space can be defined as a function of its total angular momentum $L$ and binding energy $\mathcal{E}$: $\mathcal{F} (\mathcal{E}, L; \alpha, \beta, \gamma, \Phi_0)$, an analytical expression (see \citealt{Binney2008} and \citealt{Evans1997} for details regarding the form of the distribution function). 

The implementation used here \citep{Medina:2026gme} introduces a radial normalization of the distribution function with respect to \citet{Medina2025c}, which anchors $\alpha$ to the Galactocentric distance range covered by the data, making it physically interpretable as the outer-halo tracer density slope for distances larger than a minimum radius, $r_{\rm min}$ (in contrast to previous implementations that treated $\alpha$ as a nuisance parameter).

The radial profiles of the Milky Way potential and stellar halo density are observed to have distinct breaks in their characteristic power-law slope \citep[e.g.,][]{2018ApJ...862L...1D,2022AJ....164..249H,Amarante:2024,Li:2026}.
Although a (multiply) broken power law tracer density profile \citep{Deason:2011, Deason:2013, Wang:2015, Li:2026} is in principle a better description than the single power laws assumed above, the GME measurement presented in this study is focused on distant stars (beyond 50~kpc).
In this regime, the potential and tracer density can be adequately described by a single power law \citep{Thomas:2018, Fukushima:2018, Fukushima:2019}, although caution must be exercised in extrapolation interior to that radius.

There are four model parameters $(\Phi_0,\gamma, \alpha, \beta)$ that are recovered using a hierarchical Bayesian approach that accounts for observational errors in the distances, proper motions, and line-of-sight velocities of tracers.
The distance and radial velocity uncertainties are assumed to obey single Gaussian distributions, while the proper motion uncertainties are assumed to follow a multivariate Gaussian distribution.
The fit of these parameters for the DESI DR2 data is shown in Figure~\ref{fig:gme_params}; differences in the resulting mass profiles are driven by the preferred values of $\Phi_0$ for each tracer (see \citealp{Medina:2026gme} for further discussion).

\subsection{NIMBLE} \label{sec:nimble-description}

\begin{figure}
\centering
\includegraphics[width=\linewidth]{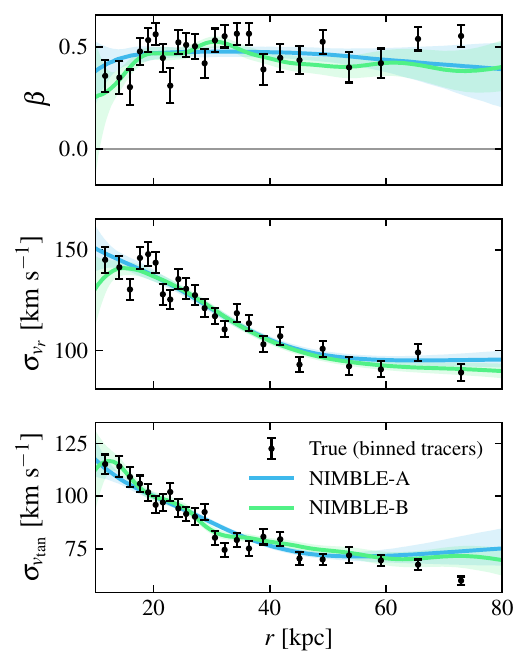}
\caption{The profiles for $\beta$, $\sigma_{v_r}$, and $\sigma_{v_\mathrm{tan}}$ from applying NIMBLE to one mock survey, compared to their true values.
The mock corresponds to the RRL catalogue for Au-6 at the 30$^\circ$ viewing.
}
\label{fig:nimble_beta}
\end{figure}

NIMBLE (Non-parametrIc jeans Modeling with B-spLinEs) solves the spherical Jeans equation, using B-splines to fit the tracer density and velocity moments.
Under the assumptions of spherical symmetry and dynamical equilibrium \citep[c.f.,][]{Binney2008},  the spherical Jeans equation relates the enclosed mass profile $M(<r)$ to the tracer density profile $\nu(r)$ and the second moments of the velocity distribution via
\begin{equation} 
    \label{sph_jeans_eq}
    M(<r) = 
    - \frac{r \, \overline{v_r^2}}{G}
    \left(
          \frac{d \ln \nu}{d \ln r} +
          \frac{d \ln \overline{v_r^2}}{d \ln r} + 
          2\beta 
    \right),
\end{equation}
where the velocity anisotropy $\beta$ is defined as
\begin{equation}
    \beta = 1 - \frac{\overline{v_\mathrm{tan}^2}}{\overline{v_r^2}}.
\end{equation}
Here $\overline{v_r^2}$, $\overline{v_\mathrm{tan}^2}$ denote the second moments of the velocity distribution, with $v_\mathrm{tan}$  the 2D tangential velocity combining the $v_\phi$ and $v_\theta$ components.

Large samples of 6D phase space coordinates for Milky Way halo stars motivate the creation of a data-driven modeling framework, using flexible B-splines to fit the observed data instead of using parametric distributions.
NIMBLE \citep{Rehemtulla2022} handles the incomplete spatial coverage and measurement uncertainties inherent to real observations with a Markov chain Monte Carlo (MCMC) algorithm that ``deconvolves'' the observational uncertainties to recover the true underlying velocity and density profiles with their statistical uncertainties in a forward-modeled Bayesian framework. 

The supporting paper by \citet{Valluri:2026nimble} describes two updated branches of the algorithm each containing different capabilities and assumptions:

\textbf{NIMBLE-A:}
\begin{enumerate}
    \item Computes the tracer density profile $\nu(r)$ using importance sampling and applies volume corrections for excised regions (namely the Sagittarius stream).
    \item Incorporates the DESI DR2 footprint and completeness function $S$ described in Section~\ref{sec:selection-functions} in the tracer density normalization.
    \item Jointly fits B-splines (within an MCMC framework) to $\nu(r)$, $\sigma_{v_r}(r)$, and $\sigma_{v_{\rm tan}}(r)$ assuming that mean streaming velocities are zero.
\end{enumerate}

\textbf{NIMBLE-B:}
\begin{enumerate}
    \item Adopts external tracer  profiles $\nu(r)$: the RRL profile from \citet{Medina2024} and the spherically averaged BHB profile from \citet{Amarante:2024}.\footnote{Despite the very different provenance of these density profiles (modeling choices, pencil-beam vs.~wide-area sky coverage), they are remarkably similar for $r>20 \kpc$, the relevant range for our data.}
    \item Fits a mixture model to the velocity distribution, using two gaussian normal components and a fixed outlier distribution to robustly manage contaminants. The gaussian normals are described by B-splines fit to $\sigma_{v_r}(r)$, $\sigma_{v_{\rm{tan}}}(r)$ and mean streaming velocities $\overline{v_r}$, $\overline{v_\phi}$, $\overline{v_\theta}$.
\end{enumerate}

Figure~\ref{fig:nimble_beta} illustrates the ability of both NIMBLE-A and NIMBLE-B to recover the underlying $\beta$, $\sigma_{v_r}$, and $\sigma_{v_\mathrm{tan}}$ profiles as a function of radius for an example AuriDESI mock.

\section{Combining the estimates} \label{sec:combination}

In this section we combine the measurements of the enclosed mass profile $M(<r)$ and total mass \Mtwoc from all independent methods and tracers.
The combined estimates on the mocks and DESI data are presented in Sections~\ref{sec:mock-results} and \ref{sec:dr2-results}, respectively.
We have posterior distributions of $M(<r)$ and \Mtwoc{} from the four methods (\empdf, GME, NIMBLE-A, NIMBLE-B) using two tracers each (RRL and BHB), resulting in $N_m=8$ estimates.
Each method has its own systematic uncertainty reflecting its sensitivity to halo structure and non-equilibrium dynamics.

We calibrate the result of each method-tracer combination using the $N=24$ AuriDESI mocks (six halos observed at four viewing angles).
We treat each combination of the halo and viewing angle as an independent mock measurement, with the caveat that different viewing angles for the same halo likely retain some correlation due to the amount of substructure and non-equilibrium.
Using the performance of each method on this finite set of mocks, we can estimate the systematic uncertainty and use it to correct the posterior distributions.
These corrections include both wider scatter and subtracting a systematic bias relative to the true values of the mocks.
Some of the supporting papers include a correction to their results based on the AuriDESI mock catalogs; we begin with the uncorrected results except those for \empdf, where we use the results after their calibration procedure that accounts for measurement uncertainties (see Section~\ref{sec:empdf-description}).

We caution that our joint estimate technique is tuned to produce unbiased results with calibrated uncertainties as applied to our mock catalogues.
This means the joint estimate, along with the computed bias and systematic uncertainties for each method, rely on the realism of those mocks.

\subsection{Enclosed mass profile} \label{sec:mass_enclosed_method}
We correct and combine mass estimates as a function of distance using a rigorous Maximum Likelihood Estimation (MLE) formalism that includes correlations between individual methods.
The full details of our methodology can be found in Appendix~\ref{app:mass-enclosed-fitting}.

We provide a joint estimate of $M(<r)$ from 20 to 80~kpc, a range that mitigates substantial contribution from the Galactic disk and does not extend beyond our tracer data.
We include the contribution of the GME method only between 50 and 80~kpc, since it only uses data within this range.

The posteriors are typically asymmetric around the median.
To capture this shape more accurately, we fit split-Gaussian distributions\footnote{Split-Gaussian distributions are created by joining two halves of two different normal distributions with a shared peak.} to the reported percentiles.
The per-method bias and covariance matrix are estimated directly from performance on the AuriDESI mocks.
Since the number of mocks (24) is insufficient to estimate all 36 elements of the covariance matrix, we apply a technique of shrinkage toward the diagonal \citep[e.g.,][]{Ledoit:2004estimator, Ledoit:2004honey} to reduce contribution of noisy off-diagonal terms.
We also apply a method-dependent floor on the systematic uncertainty, which helps balance the effective contributions of the individual methods to the combined estimate (avoiding a joint estimate that is effectively driven by one method).

Before applying this method to the DESI data, we validate it on the mocks using a leave-one-out procedure.
For each mock $i$, calibration is performed on the remaining $N-1$ mocks, and the combination is applied to mock $i$.
The shrinkage parameter $\lambda$ and the per-method floor $f_m$ are optimized such that the coverage of the 68~per~cent interval is closest to 68~per~cent and the standard deviation of the error-weighted distribution of residuals is closest to unity.
Given only 24 mocks, the best calibration achieves 70.8~per~cent coverage (slight overestimation of errors) and the standard deviation of 1.002.
The mean residual is only +0.0014~dex, consistent with zero.
The optimized values of $\lambda$ and $f_m$ are consistent for all distances between 20 and 80~kpc.
We also validated the algorithm on simple examples with known analytical answers.
This test confirms that the algorithm works correctly and produces accurate error estimates.

No single method-tracer combination contributes more than 40~per~cent weight on the joint estimate at a given distance, and often the maximum weight is near 30~per~cent.
The method that contributes the most weight (summed across tracers), either NIMBLE-A or NIMBLE-B, stays within 40 and 70~per~cent across all distances.
RRLs and BHBs contribute nearly equally to the joint estimate, with both weights (summed across methods) staying within 40 and 60~per~cent across all distances.

\subsection{Total mass}

Estimating the total mass \Mtwoc{} of the Galaxy from our dataset requires significant extrapolation to large distances.
In principle, we could fit an assumed profile to the joint enclosed mass estimate $M(<r)$ and then extrapolate this profile to produce a joint \Mtwoc.
However, this would combine the different methods' assumptions, biases, and errors in a complex way.
Instead, we derive joint estimates by combining the \Mtwoc{} values reported by each method.

The \empdf{} method already assumes an NFW mass distribution and reports \Mtwoc{} as a fit parameter.
GME and NIMBLE estimate \Mtwoc{} by fitting an NFW profile and fixed baryonic mass profile to their enclosed mass constraint.
Both adopt the baryonic mass distribution from \texttt{MilkyWayPotential2022} implemented in \texttt{gala} \citep{gala}.
For GME, the NFW profile is fit to the enclosed mass and power law slope estimated at $r = 65\kpc$, halfway along the radial range considered.
For both NIMBLE methods, the profile is fit to the enclosed mass across a broad radial range.

We take a broadly similar approach to combining \Mtwoc estimates as the joint $M(<r)$, with some implementation differences.
Rather than the MLE point-calibration used for $M(<r)$, we fit a hierarchical Bayesian model in which the per-method biases, floors, error scalings, and correlation matrix are given priors and sampled.
The calibration uncertainty is then included into the final posterior.
The full details can be found in Appendix~\ref{App:M200Fitting}.
It is worth mentioning that the \Mtwoc masses are constrained jointly with a free concentration parameter ($c_\text{200c}$), such that the uncertainties are at least two-dimensional.
The mass uncertainties quoted here follow from marginalising over concentration.
Because mass and concentration are covariant, the joint confidence contour is elongated along their degeneracy direction, and at a fixed confidence level it reaches further in mass than the corresponding marginalised interval.

Similar to the joint $M(<r)$, no single method-tracer combination contributes more than 40~per~cent of the weight for the joint \Mtwoc.
RRLs and BHBs contribute nearly equally.
Summed across tracers, the GME method contributes 75~per~cent of the weight, driven by its accurate and precise \Mtwoc estimates for the mocks (see Section~\ref{sec:m200-mock-results}).

\section{Results on the AuriDESI mocks} \label{sec:mock-results}

\begin{figure}
\includegraphics[width=\linewidth]{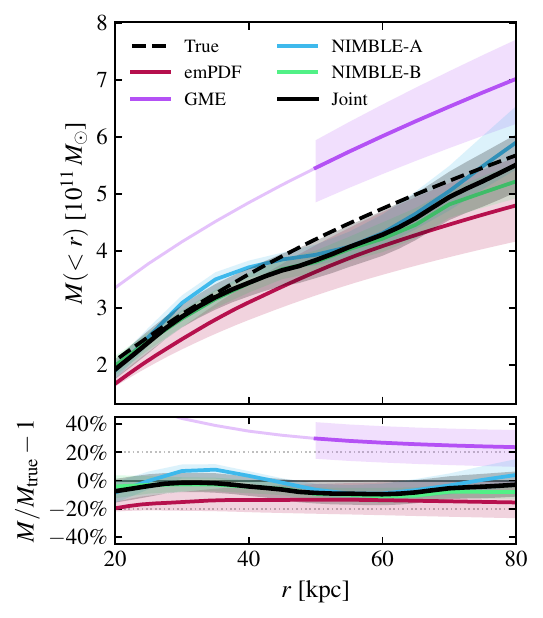}
\caption{Example comparison between the four individual methods, along with the joint estimate, for one mock survey.
The mock corresponds to the RRL catalogue for Au-6 at the 30$^\circ$ viewing.
We present similar results for all mocks in Appendix~\ref{app:all-mocks-figures}.
}
\label{fig:menc-au6-comparison}
\end{figure}

\begin{figure*}
\centering
\includegraphics[width=\linewidth]{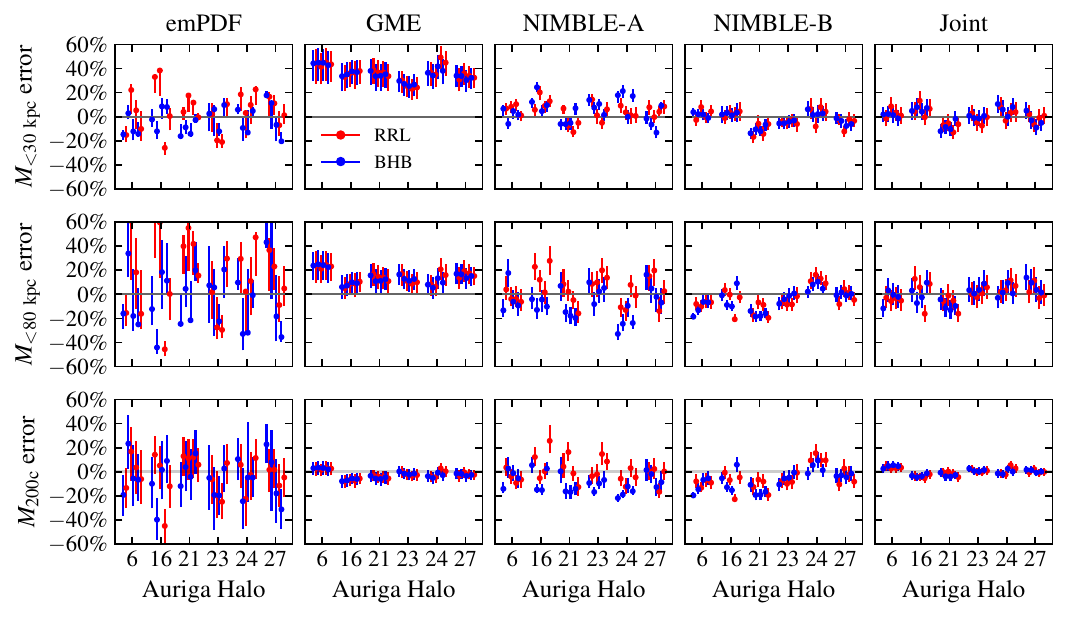}
\caption{Performance of the individual methods on all AuriDESI mock catalogues.
The mass estimates are shown for $M(r<30 \kpc)$, $M(r<80 \kpc)$, and \Mtwoc in the top, middle, and bottom panels, respectively.
Each point shows the median value, while the error bar is the 16-84~per~cent scatter.
}
\label{fig:rel-menc-comparison}
\end{figure*}

We now compare the results from the different modeling techniques applied to the AuriDESI mock catalogues.
Here, the results we show for individual methods are not corrected for the biases determined in Section~\ref{sec:combination}, but the joint estimate includes these corrections.

\subsection{Enclosed mass profile}
As an example of variation in mass profile recovery, in Figure~\ref{fig:menc-au6-comparison} we show the results from applying all methods to the same mock catalog (the RRL mock for Au-6 at the 30$^\circ$ viewing angle).
The dashed curve shows the true enclosed mass profile of the simulated halo, computed by summing the masses of star, dark matter, and gas particles in spherical radial bins. 
The lower panels show the percentage error on the enclosed mass as a function of radius.
For this particular mock, the GME estimate is biased high ($\sim$20-30~per~cent), the other estimates fall within $\sim$15-20~per~cent of the truth, and the joint estimate is well calibrated ($\lesssim$10~per~cent) across all distances.
Similar enclosed mass profiles for all mocks can be seen in Appendix~\ref{app:all-mocks-figures}, for both RRLs (Figure~\ref{fig:menc-comparison-all-rrl}) and BHBs (Figure~\ref{fig:menc-comparison-all-bhb}).

To illustrate the variance across all haloes, in Figure~\ref{fig:rel-menc-comparison} we show the relative errors across all mock catalogues for three example quantities: $M(<30\kpc)$, $M(<80\kpc)$, and \Mtwoc.
Comfortingly, it is difficult to identify any trends for a given halo that are true across all methods.
For example, at a given distance, the halo with the worst accuracy (largest mean offset) is never the same across all four individual methods.
The same is true for the halo with the largest spread in mass estimates across viewing angles.
This suggests that halo-specific features, like amount of substructure or dynamical disequilibrium, do not affect all methods the same way.
However, we note some method combinations can be correlated across different views of the same halo (e.g., \empdf and NIMBLE-A for the estimate of $M(<80\kpc)$ in Au-27).
We discuss these correlations in greater detail in Section~\ref{sec:correlations}.

\begin{figure}
\centering
\includegraphics[width=\linewidth]{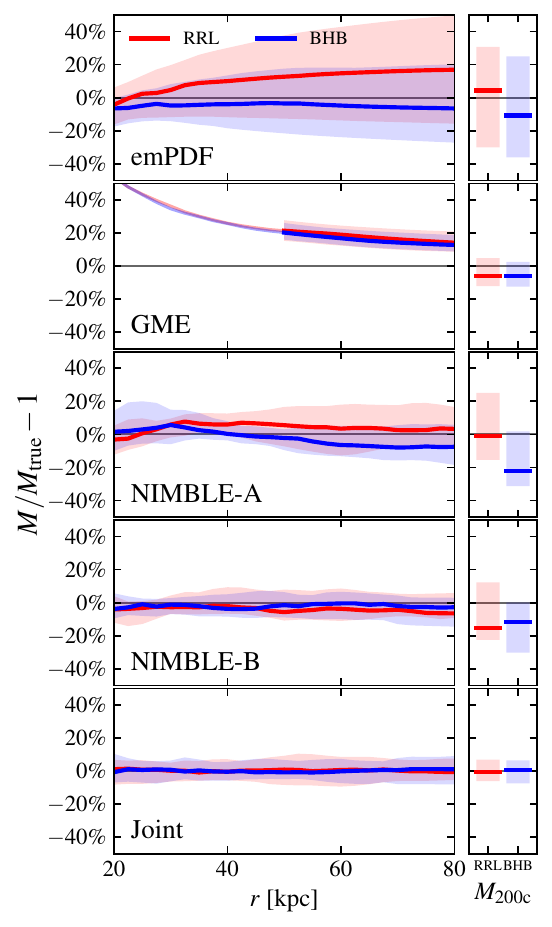}
\caption{Median bias and scatter of each method across all AuriDESI mock catalogues.
Each row corresponds to an individual method, with the joint estimate at the bottom. 
The left panels show the bias for the enclosed mass $M(<r)$ and the right panels show the bias for \Mtwoc.
The GME method bias is only shown from 50\kpc outwards.
}
\label{fig:method-auridesi-summary}
\end{figure}

\begin{deluxetable}{llcc}
\tablecaption{Table of biases and scatter for $M(<r)$ and \Mtwoc based on AuriDESI mock tests (as displayed in Figure~\ref{fig:method-auridesi-summary}).
The values for $M(<r)$ are the medians across the considered radial range (from 20 to 80~kpc).
Values are reported as percentages.}
\label{tab:mock_accuracy_summary}
\tablehead{
\colhead{Method} &
\colhead{Tracer} &
\colhead{$M(<r)$} &
\colhead{\Mtwoc}
}
\startdata
emPDF & RRL & $+12.7 \pm 25.2$ & $+4.3 \pm 30.5$ \\
 & BHB & $-4.8 \pm 19.6$ & $-10.8 \pm 30.5$ \\
GME & RRL & $+21.4 \pm 6.2$ & $-6.2 \pm 8.5$ \\
 & BHB & $+20.2 \pm 5.0$ & $-5.9 \pm 7.4$ \\
NIMBLE-A & RRL & $+3.9 \pm 11.9$ & $-1.2 \pm 20.3$ \\
 & BHB & $-2.2 \pm 9.5$ & $-22.1 \pm 16.6$ \\
NIMBLE-B & RRL & $-4.1 \pm 7.8$ & $-15.3 \pm 17.4$ \\
 & BHB & $-2.1 \pm 7.7$ & $-11.6 \pm 15.2$ \\
\hline
Joint & RRL & $+0.3 \pm 6.7$ & $-0.4 \pm 6.5$ \\
 & BHB & $+0.1 \pm 6.6$ & $+0.5 \pm 6.9$ \\
 & RRL + BHB & $-0.0 \pm 6.0$ & $+0.6 \pm 7.5$ \\
\enddata
\end{deluxetable}

In Figure~\ref{fig:method-auridesi-summary} we show the trends of $M(<r)$ offsets averaged across the entirety of the mock sample.
This allows the behavior of the individual methods to become apparent, over the mock-to-mock scatter.
The median bias and scatter values from the 20 to 80~kpc range for all of the methods are summarised in Table~\ref{tab:mock_accuracy_summary}.

\empdf has broadest scatter with relatively small overall bias, with the BHBs achieving better than 5~per~cent offset across the full distance range of the data.
\empdf also tends to report larger formal uncertainties than other methods (Figure~\ref{fig:rel-menc-comparison}).
However, both our joint estimate and \citet{Zhou:2026empdf} find that the scatter across all available mocks is slightly larger than the reported formal uncertainties (derived in slightly different ways).

For GME, mass estimates at $r < 50 \kpc$ are extrapolations of the inferred enclosed mass profile.
Figure~\ref{fig:rel-menc-comparison} shows that the estimated $M(<r)$ profiles are remarkably consistent across the two tracer populations for the same mock halo and viewing. 
However, GME consistently overestimates the enclosed mass by $\sim$20~per~cent for all mocks within $50 \kpc < r < 80 \kpc$, with a small scatter of 5-6~per~cent.

On average across the sample, both NIMBLE-A and NIMBLE-B recover the mass enclosed to $\sim$10~per~cent at all radii.
We note that for these tests, NIMBLE-A reconstructs the number density profile for each mock viewing, while NIMBLE-B uses the same spherically-averaged density profile for each halo.
This ``external tracer density profile'' is computed for star particles in the Auriga halo with [Fe/H]~$<-1.5$ and age~$>8$~Gyr. 
NIMBLE-A shows a slight overbias, which is stronger for the RRL and is likely driven by the accuracy of the reconstructed density profile.
NIMBLE-B shows a small under bias and slightly smaller scatter ($\sim$8~per~cent) relative to NIMBLE-A ($\sim$10~per~cent), which may be optimistic since the density profile is computed using true simulation data.
In Figure~\ref{fig:rel-menc-comparison} we see strong systematic effects in NIMBLE across different viewings of the same halo, likely resulting from substructure and disequilibrium.

The joint estimates have essentially no bias across the entire radial range (by construction), with the smallest scatter of 6 to 7~per~cent.
This is also true for individual haloes, as seen in Figure~\ref{fig:rel-menc-comparison}.
In contrast, the individual estimates from the four methods can have significant outliers, even after subtracting their overall bias.
This confirms the reliable performance of the joint estimate.

\subsection{Total mass} \label{sec:m200-mock-results}

\begin{figure*}
    \includegraphics[width=\linewidth]{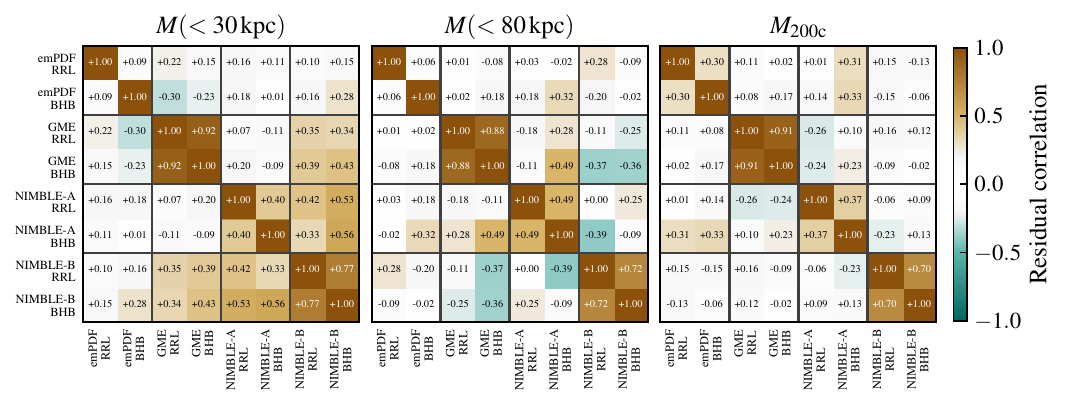}
    \caption{
    The correlation matrix of the eight method--tracer pairs, as measured from their residuals (i.e., after subtracting the relevant true quantity) on the 24 AuriDESI mocks.
    Correlations smaller than $\pm 0.2$ should be considered too small to be meaningful with our limited sample of mocks.
    }
    \label{fig:corr_matrix}
\end{figure*}

Similarly to the enclosed mass profiles, we also assess the capability of the four methods and the joint technique to recover \Mtwoc using the AuriDESI mocks.
We show the per-mock performance in the bottom panel of Figure~\ref{fig:rel-menc-comparison}, the averaged performance across all mocks in the right panels of Figure~\ref{fig:method-auridesi-summary}, and the correlation between methods in the right panel of Figure~\ref{fig:corr_matrix}.

Broadly, the trends reported in the enclosed mass estimates hold for the \Mtwoc results as well.
The results reported by \empdf are relatively unbiased but have the largest scatter and uncertainties, with similar performance across all individual mocks.
This is consistent with the minimal assumption nature of \empdf.
Both NIMBLE methods show wider scatter in \Mtwoc{} than they did for the enclosed mass range, mostly reflecting the uncertainty of extrapolation.
Perhaps most interesting, GME has a very small underbias and scatter for \Mtwoc, much closer to the truth than the enclosed mass profile.
This could mean that the assumptions baked into the GME method, combined with our extrapolation procedure, are more justified when estimating the total mass of a halo, rather than its profile.

Similarly to the joint enclosed mass, the joint \Mtwoc estimates have essentially no bias and the smallest scatter of 6 to 7~per~cent.
We emphasize that the joint \Mtwoc is not simply copying the bias-corrected GME results; the per-method error floor and additional systematics (see Appendix~\ref{App:M200Fitting}) help guard against a single method dominating the final estimate, though we note that GME is assigned $\sim$75~per~cent of the \Mtwoc weight.

\subsection{Residual correlations between methods} \label{sec:correlations}

Both the joint $M(<r)$ and \Mtwoc methodologies account for the residual correlations between individual methods when combining the estimates (see Section~\ref{sec:combination} and Appendix~\ref{app:combined-math-both} for details).
These correlations should be interpreted as an empirical determination of combined statistical correlations (e.g., using the same mock dataset or DESI catalogs) and systematic correlations (e.g., halo substructure and disequilibrium).
In Figure~\ref{fig:corr_matrix} we show these correlations for $M(<30\kpc)$, $M(<80\kpc)$, and \Mtwoc.

The residual correlations vary as a function of distance, with nearby distances showing the strongest maximum correlations across all method-tracer pairs.
Some combinations can even change from being positively correlated at small distance to negatively correlated at large distance.
We caution that due to our relatively small sample of mocks (24 for a given tracer), correlations within $\pm 0.2$ should be considered too small to be meaningful.

For a given mass quantity, the strongest correlations are usually found when applying the same method using different tracers.
However, this is not true for all methods; \empdf and NIMBLE-A both show relatively low residual correlations between their RRL and BHB results.
The relatively low correlations between NIMBLE-A and NIMBLE-B supports their treatment as separate measurements, even though they share many assumptions and implementation details.
Much of the strength of the joint estimates shown in Figures~\ref{fig:rel-menc-comparison} and \ref{fig:method-auridesi-summary} comes from modeling these correlations in a robust manner, enabled by benchmarking on the same set of mocks.

\section{Results on DESI DR2} \label{sec:dr2-results}

We now present the results for the individual methods and joint estimate for the DESI DR2 Milky Way data.
Here, the results we show for individual methods are corrected for the biases determined in Section~\ref{sec:combination}.
The joint estimates additionally account for correlations between the individual estimates.

\subsection{Enclosed mass profile of the Milky Way}

\begin{figure*}
\includegraphics[width=\linewidth]{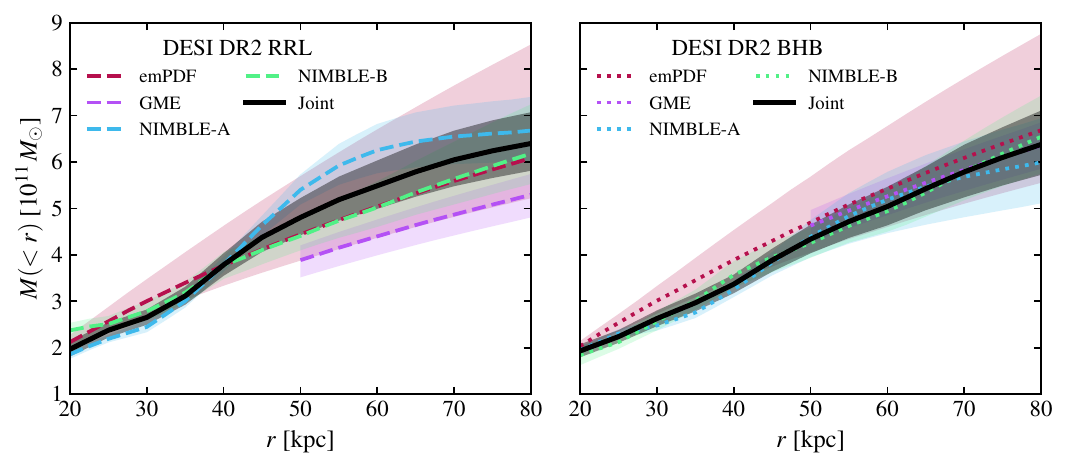}
\caption{Joint constraints on the Milky Way mass profile and the individual method estimates, separately for the RRL (left) and BHB (right) tracers.
Shaded regions show the 68~per~cent confidence intervals.
The results for individual methods are corrected for their biases, following the procedure in Section~\ref{sec:mass_enclosed_method}.
}
\label{fig:mw_mass_separate}
\end{figure*}

\begin{figure}
\includegraphics[width=\linewidth]{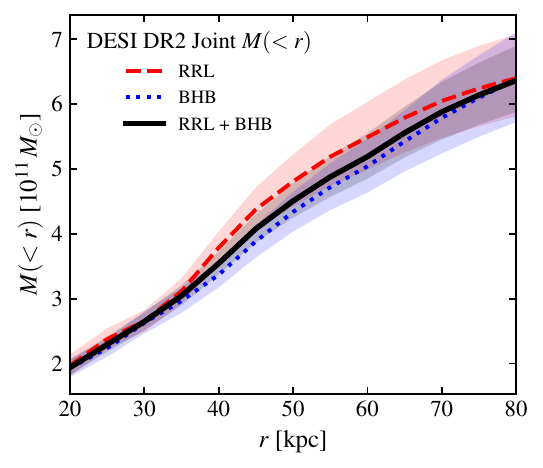}
\caption{Joint DESI DR2 constraints on the Milky Way mass profile using only RRL-based estimates (dashed), only BHB-based estimates (dotted), and combining all estimates (solid).
}
\label{fig:mw_mass_joint}
\end{figure}

\begin{deluxetable*}{ccccccc}
\tablecaption{Joint estimate of the Milky Way mass profile, combining both tracers and using them separately.}
\label{tab:mw_mass_profile_rrl_bhb}
\tablehead{
\colhead{$r$} &
\colhead{$\log_{10}(M_{\rm MW}/M_\odot)_{\rm All}$} &
\colhead{$M_{\rm MW,All}$} &
\colhead{$\log_{10}(M_{\rm MW}/M_\odot)_{\rm RRL}$} &
\colhead{$M_{\rm MW,RRL}$} &
\colhead{$\log_{10}(M_{\rm MW}/M_\odot)_{\rm BHB}$} &
\colhead{$M_{\rm MW,BHB}$} \\
\colhead{[kpc]} &
\colhead{} &
\colhead{$[10^{11}\,M_\odot]$} &
\colhead{} &
\colhead{$[10^{11}\,M_\odot]$} &
\colhead{} &
\colhead{$[10^{11}\,M_\odot]$}
}
\startdata
20 & $11.287^{+0.026}_{-0.027}$ & $1.94^{+0.12}_{-0.12}$ & $11.294^{+0.037}_{-0.037}$ & $1.97^{+0.17}_{-0.16}$ & $11.286^{+0.031}_{-0.032}$ & $1.93^{+0.14}_{-0.14}$ \\[2mm]
25 & $11.361^{+0.022}_{-0.022}$ & $2.30^{+0.12}_{-0.11}$ & $11.376^{+0.030}_{-0.030}$ & $2.38^{+0.17}_{-0.16}$ & $11.350^{+0.027}_{-0.027}$ & $2.24^{+0.14}_{-0.13}$ \\[2mm]
30 & $11.423^{+0.022}_{-0.022}$ & $2.65^{+0.14}_{-0.13}$ & $11.424^{+0.027}_{-0.027}$ & $2.65^{+0.17}_{-0.16}$ & $11.420^{+0.028}_{-0.028}$ & $2.63^{+0.18}_{-0.16}$ \\[2mm]
35 & $11.484^{+0.022}_{-0.022}$ & $3.05^{+0.16}_{-0.15}$ & $11.493^{+0.027}_{-0.027}$ & $3.11^{+0.20}_{-0.19}$ & $11.472^{+0.028}_{-0.028}$ & $2.97^{+0.20}_{-0.18}$ \\[2mm]
40 & $11.549^{+0.021}_{-0.021}$ & $3.54^{+0.18}_{-0.17}$ & $11.578^{+0.028}_{-0.028}$ & $3.79^{+0.25}_{-0.24}$ & $11.527^{+0.026}_{-0.026}$ & $3.37^{+0.21}_{-0.19}$ \\[2mm]
45 & $11.610^{+0.023}_{-0.023}$ & $4.08^{+0.22}_{-0.21}$ & $11.641^{+0.032}_{-0.032}$ & $4.38^{+0.34}_{-0.31}$ & $11.589^{+0.027}_{-0.029}$ & $3.89^{+0.25}_{-0.25}$ \\[2mm]
50 & $11.654^{+0.025}_{-0.026}$ & $4.51^{+0.27}_{-0.26}$ & $11.682^{+0.036}_{-0.036}$ & $4.81^{+0.42}_{-0.38}$ & $11.637^{+0.030}_{-0.032}$ & $4.34^{+0.31}_{-0.30}$ \\[2mm]
55 & $11.688^{+0.028}_{-0.028}$ & $4.88^{+0.32}_{-0.31}$ & $11.715^{+0.040}_{-0.040}$ & $5.19^{+0.50}_{-0.45}$ & $11.674^{+0.033}_{-0.034}$ & $4.72^{+0.37}_{-0.35}$ \\[2mm]
60 & $11.715^{+0.029}_{-0.029}$ & $5.19^{+0.36}_{-0.34}$ & $11.739^{+0.041}_{-0.041}$ & $5.49^{+0.55}_{-0.50}$ & $11.703^{+0.035}_{-0.036}$ & $5.04^{+0.42}_{-0.40}$ \\[2mm]
65 & $11.745^{+0.030}_{-0.031}$ & $5.56^{+0.40}_{-0.38}$ & $11.763^{+0.042}_{-0.042}$ & $5.79^{+0.59}_{-0.54}$ & $11.734^{+0.037}_{-0.039}$ & $5.43^{+0.48}_{-0.46}$ \\[2mm]
70 & $11.769^{+0.033}_{-0.033}$ & $5.88^{+0.46}_{-0.43}$ & $11.782^{+0.043}_{-0.043}$ & $6.05^{+0.63}_{-0.57}$ & $11.762^{+0.042}_{-0.043}$ & $5.79^{+0.59}_{-0.55}$ \\[2mm]
75 & $11.788^{+0.034}_{-0.034}$ & $6.14^{+0.50}_{-0.46}$ & $11.795^{+0.043}_{-0.043}$ & $6.24^{+0.66}_{-0.59}$ & $11.785^{+0.044}_{-0.045}$ & $6.10^{+0.65}_{-0.60}$ \\[2mm]
80 & $11.804^{+0.035}_{-0.034}$ & $6.36^{+0.53}_{-0.48}$ & $11.806^{+0.043}_{-0.042}$ & $6.40^{+0.67}_{-0.59}$ & $11.805^{+0.047}_{-0.047}$ & $6.38^{+0.73}_{-0.66}$ \\[2mm]
\enddata
\end{deluxetable*}

The enclosed mass profiles from applying our individual methods to the DESI DR2 RRL and BHB catalogues are shown in Figure~\ref{fig:mw_mass_separate}.
We also show the equivalent joint profiles that combine only the RRL results and only the BHB results.
These joint profiles are in broad agreement with each other and fall within the span of individual estimates for a given tracer.
However, where individual methods fall relative to this span (and the joint estimate) differ between the tracers.

For \empdf, the two tracer populations give consistent results over the radial range constrained by the data.
As \empdf{} directly fits $(M_{200\mathrm{c}}$ and $c_{200\mathrm{c}})$ assuming an NFW profile, the posterior contours for these two parameters are its primary constraint.
The resulting $M(<r)$ profiles have nearly identical $1\sigma$ uncertainty regions, with BHBs giving a slightly higher median value for $r > 50 \kpc$.

For GME, the RRL and BHB $M(<r)$ results are in modest tension, with the RRL result having a lower mass.
This is notable, as for the AuriDESI mocks the GME results were always very close between the two tracers.
\citet{Medina:2026gme} find that the BHB result is somewhat sensitive to a small number of stars with large tangential velocities ($v_\text{tan} > 500$km~s$^{-1}$), though most of these stars are removed by the \texttt{RVS\_WARN = 0} selection in the fiducial sample.

The NIMBLE-B estimates for BHBs and RRLs are in remarkable agreement with each other across nearly all radii.
The NIMBLE-A results agree with each other within 30~kpc, but beyond this distance the RRL result rises much more quickly and flattens at a much higher mass than the BHB result (but similar to the NIMBLE-B result).
At 80~kpc, the two have overlapping $1\sigma$ uncertainty regions but the median BHB estimate is slightly smaller than the 16th percentile for the RRL estimate.
This is likely due to NIMBLE-A's inferred RRL number density profile declining more steeply than the profile adopted by NIMBLE-B (see \citealp{Valluri:2026nimble} for details).

The combined $M(<r)$ estimate using all methods and both tracers is presented in Figure~\ref{fig:mw_mass_joint}, along with the per-tracer joint estimates from Figure~\ref{fig:mw_mass_separate}.
Despite significant variation between the individual methods, the joint estimate for all tracers increases steadily with Galactocentric radius.
It lies in the middle of the most discrepant estimates, and is closest to the \empdf and NIMBLE-B estimates.
The joint RRL profile rises slightly faster than the other two joint estimates, mostly reflecting the rapid rise in both NIMBLE profiles from 30 to 40~kpc and shallower slope for NIMBLE-B beyond this distance (due to a lower and more steeply falling $\sigma_r$).
Table~\ref{tab:mw_mass_profile_rrl_bhb} provides all three joint estimates in 5 kpc intervals.
The joint estimate uncertainty is under 8~per~cent across the full range of distances, roughly half that of the tightest individual method.
The mass estimate at the largest distance directly probed by DESI DR2 is $M(r < 80\kpc) = \JointMencEighty$.

\subsection{Total mass of the Milky Way}

\begin{figure}
\centering
\includegraphics[width=\linewidth]{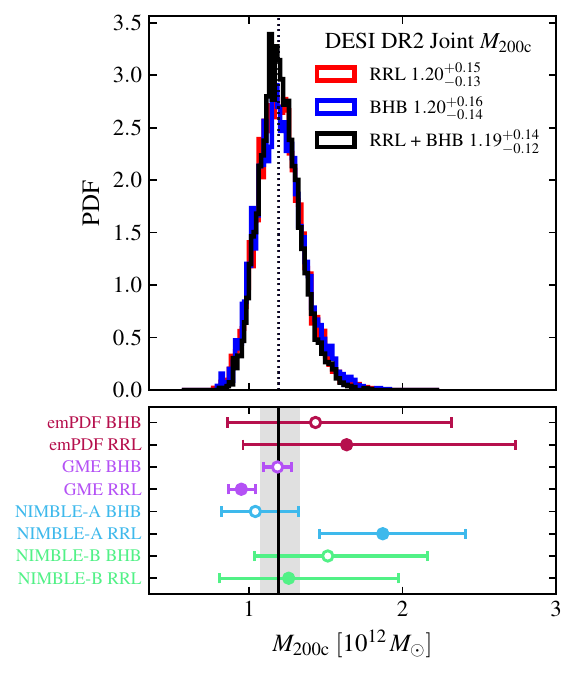}
\caption{Posterior distribution of the total mass \Mtwoc{} from the joint estimate (top panel) and corrected individual estimates (bottom panel).
We also show separate results combining only RRL measurements (red) and only BHB measurements (blue).
}
\label{fig:m200_pdf}
\end{figure}

\begin{deluxetable}{llcc}
\tablecaption{Calibrated \Mtwoc of the Milky Way for each method and tracer, along with the resulting joint RRL, BHB, and total estimates.
Note that the individual results are corrected using the joint methodology described in Section~\ref{sec:combination}; for each method's own reported results, see the supporting papers.}
\label{tab:m200_methods}
\tablehead{
\colhead{Method} &
\colhead{Tracer} &
\colhead{$\log_{10} ( \Mtwoc / M_\odot)$} &
\colhead{\Mtwoc} \\
\colhead{} &
\colhead{} &
\colhead{} &
\colhead{$[10^{12}\,M_\odot]$}
}
\startdata
\empdf & RRL & $12.214_{-0.231}^{+0.224}$ & $1.64_{-0.67}^{+1.10}$ \\ [1mm]
 & BHB & $12.156_{-0.220}^{+0.209}$ & $1.43_{-0.57}^{+0.89}$ \\ [1mm]
GME & RRL & $11.978_{-0.040}^{+0.040}$ & $0.95_{-0.08}^{+0.09}$ \\ [1mm]
 & BHB & $12.074_{-0.034}^{+0.032}$ & $1.19_{-0.09}^{+0.09}$ \\ [1mm]
NIMBLE-A & RRL & $12.272_{-0.109}^{+0.109}$ & $1.87_{-0.42}^{+0.54}$ \\ [1mm]
 & BHB & $12.018_{-0.103}^{+0.104}$ & $1.04_{-0.22}^{+0.28}$ \\ [1mm]
NIMBLE-B & RRL & $12.100_{-0.192}^{+0.196}$ & $1.26_{-0.45}^{+0.72}$ \\ [1mm]
 & BHB & $12.180_{-0.166}^{+0.155}$ & $1.51_{-0.48}^{+0.65}$ \\ [1mm]
\hline
Joint & RRL & $12.078_{-0.052}^{+0.051}$ & $1.20_{-0.13}^{+0.15}$ \\ [1mm]
 & BHB & $12.078_{-0.054}^{+0.054}$ & $1.20_{-0.14}^{+0.16}$ \\ [1mm]
 & RRL + BHB & $12.076_{-0.046}^{+0.047}$ & $1.19_{-0.12}^{+0.14}$ \\ [1mm]
\enddata
\end{deluxetable}

The DESI DR2 Milky Way \Mtwoc measurements are presented in Figure~\ref{fig:m200_pdf} and Table~\ref{tab:m200_methods}.
We find a joint estimate of $\Mtwoc = \JointMtwoc$ when combining all methods and tracers, $\JointMtwocRRL$ when combining only RRL measurements, and $\JointMtwocBHB$ when combining only BHB measurements.
This exceptional agreement among the three joint estimates is not driven by a single methodology or tracer.
 
We also report the corrected individual estimates for each method-tracer combination.
These individual measurements are broadly consistent with the joint estimate, with the exception of the RRL estimates for GME and NIMBLE-A.
The GME RRL estimate disagrees with the joint estimates (and the BHB RRL estimate) at the $\sim$2$\sigma$ level, given their very small formal errors.
The NIMBLE-A RRL estimate is somewhat high, but is consistent with the RRL estimates from both \empdf and NIMBLE-B.
There is no preference across all methods for one tracer to yield higher total mass estimates than the other.

We note that our joint estimates have a substantial weight ($\sim$75~per~cent) from the GME method, due to its remarkable performance on \Mtwoc for our mocks (see Section~\ref{sec:m200-mock-results}).
In particular, the relatively small uncertainties on our joint estimates are largely driven by this weight.
However, assigning a high \Mtwoc weight to a method with significant measurement bias within 80~kpc (Figure~\ref{fig:method-auridesi-summary}) may need to be treated with some caution.
If GME is excluded entirely from the joint analysis, we obtain a mass estimate (across all remaining methods and both tracers) of $\Mtwoc = \JointMtwocNoGME$, inflating the relative uncertainty from $\sim$11 to $\sim$22 per cent.
This represents the most extreme case of ignoring all information from GME, rather than adjusting our methodology for some intermediate outcome (e.g., introducing a maximum weight, or using weights determined for the enclosed mass profile).

\section{Discussion} \label{sec:discussion}

In this Section we compare our joint $M(<r)$ and \Mtwoc constraints to those from prior literature (Section~\ref{sec:prior-constraints}).
We discuss the growing tension between Galactic halo and rotation curve measurements (Section~\ref{sec:rotcurve-tension}) and whether Milky Way mass measurements are in the systematics-dominated regime (Section~\ref{sec:systematics-regime}).
We also comment on the relationship between our joint estimate and the individual measurements that underpin it (Section~\ref{sec:relationship-individual-joint}).

\subsection{Comparison to prior constraints} \label{sec:prior-constraints}

\begin{figure*}
\includegraphics[width=\linewidth]{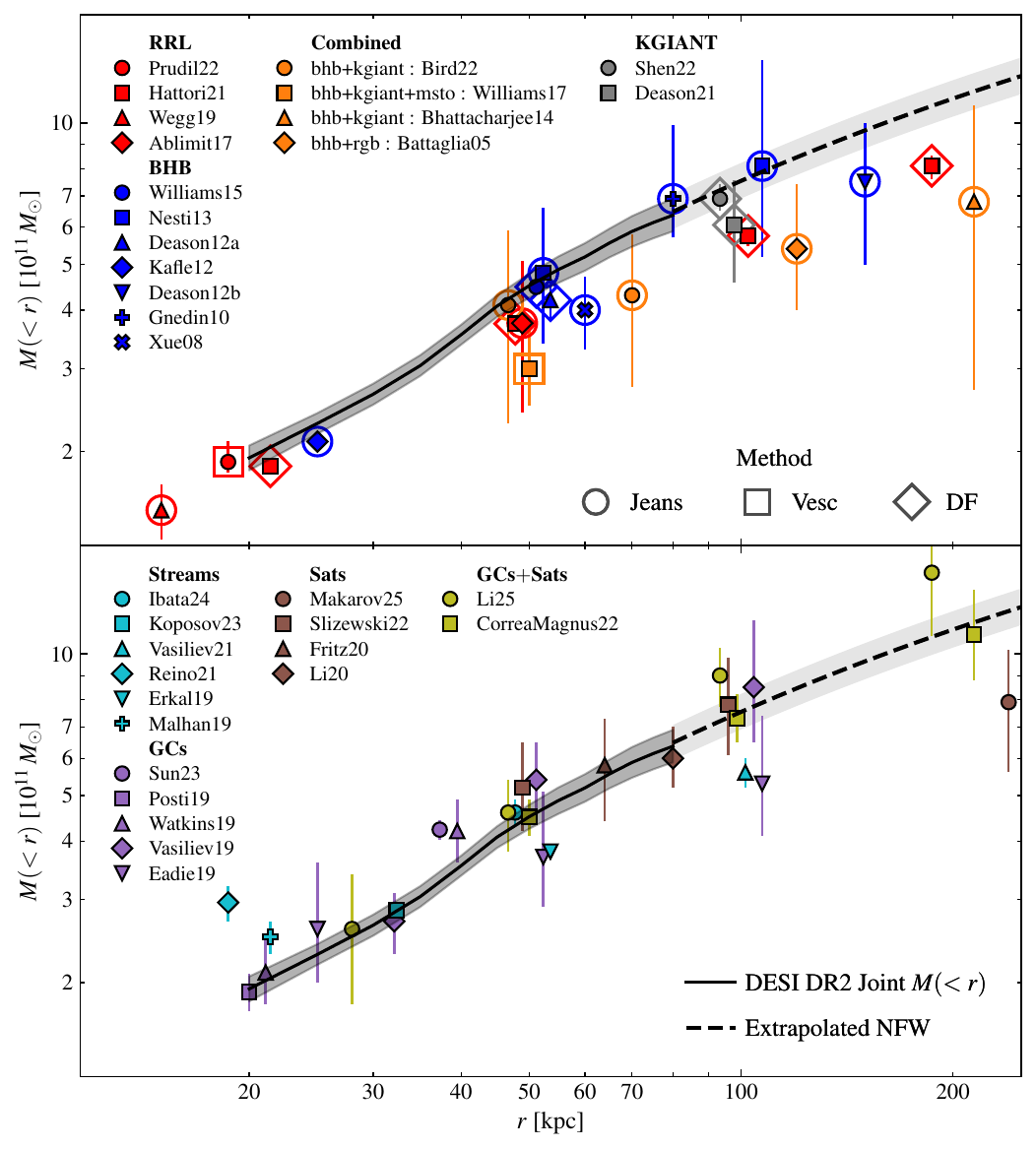}
\caption{A comparison of the DESI DR2 joint $M(<r)$ profile (solid) to results from the literature.
We also show an extrapolated NFW profile (dashed), adopting the joint \Mtwoc constraint and a concentration of $c_\text{200c} = 12$ (see text for details).
\textit{Top Panel:} The results of methods applied to field halo stars, with different colours representing the stellar population (RRL, BHB, K giant, and combined) and outer shapes indicating the methodology (spherical Jeans equations, escape velocity, and distribution function).
\textit{Bottom Panel:} The results of methods applied to different tracers, including stellar streams, globular clusters, and satellite galaxies.
For brevity, we limit results to no earlier than 2019, reflecting improved data quality since \textit{Gaia}~DR2.
Measurements reported at the same distance are displaced slightly to improve legibility.
References are listed in Appendix~\ref{app:lit-compilation-references}.
}
\label{fig:MencLit}
\end{figure*}

\begin{figure*}
\includegraphics[width=\linewidth]{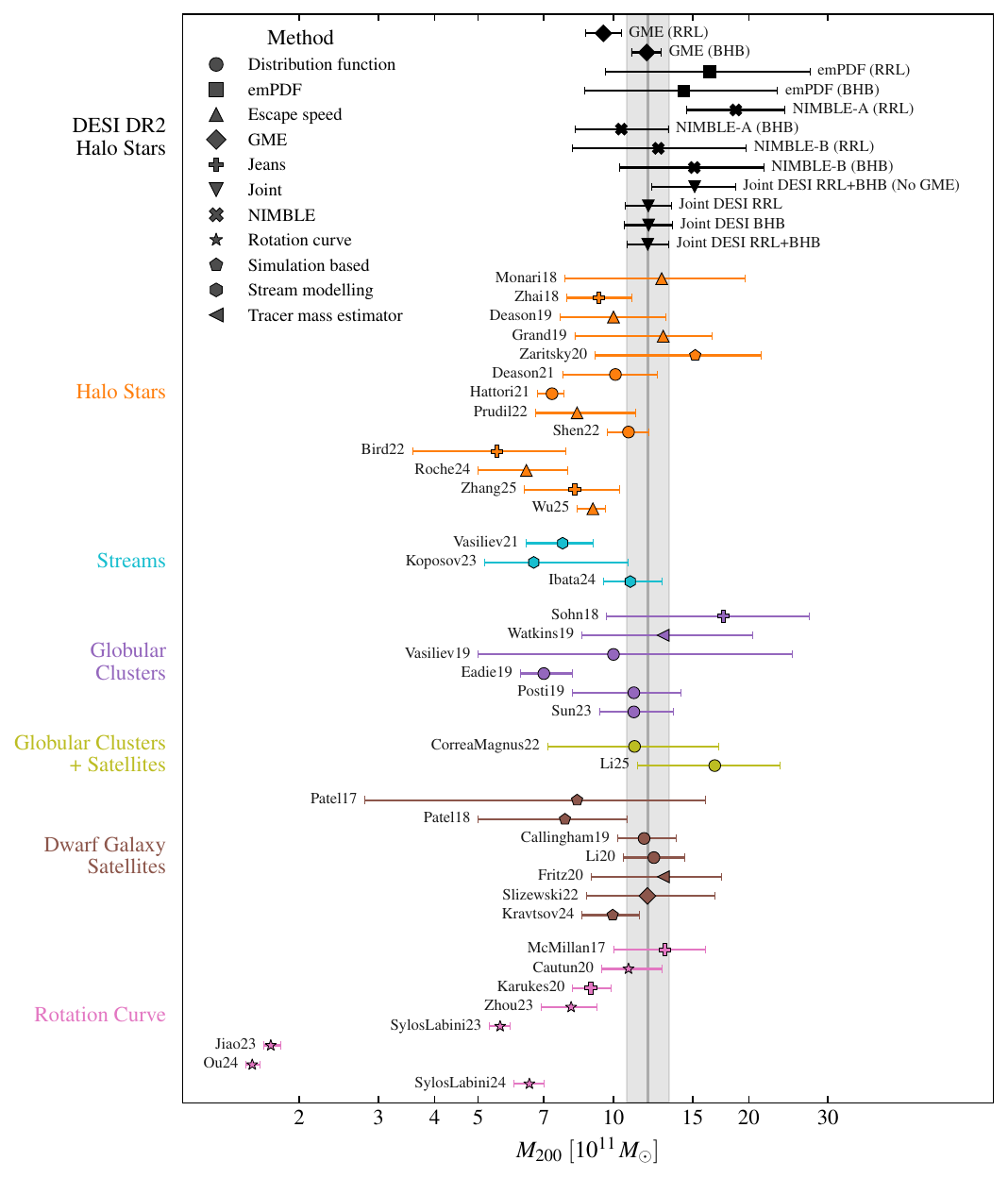}
\caption{Comparison of the $M_\text{200c}$ measurements from this work to literature values.
We include the estimates from our four individual methods using either RRL or BHB tracers, as well as the joint RRL, BHB, and combined estimates.
The joint combined estimate is also shown with a vertical line and shaded region.
The literature results are grouped and coloured by the tracer used, while the central marker of each result denotes the methodology.
References are listed in Appendix~\ref{app:lit-compilation-references}.
}
\label{fig:m200-lit-comparison}
\end{figure*}

In Figure~\ref{fig:MencLit} we present a compilation of enclosed mass estimates from the literature, compared to the DESI DR2 joint $M(<r)$ profile.
In general, our joint result is in very good agreement with the wider literature.
In the top panel, we compare to other estimates using field halo stars.
We see no clear trends in results from different stellar populations, including RRL and BHB as used in this work.
We also see no clear differences across broad categories of methodology.
Analyses based on modeling Jeans equations or distribution functions typically rely on large ensembles of stars, while methods that infer escape velocities typically place a greater weight on a small number of high-velocity stars.

In the bottom panel, we see that results from other tracers are in broad agreement with our measurement as well.
The stellar streams used for mass modeling mostly lie within 50~kpc, with some methods providing additional constraints on the shape of the Galactic mass distribution, which we do not model and do not indicate on these plots.
The globular clusters occupy a similar radial range to the field halo stars, with fewer individual tracers but often better quality data for each point tracer.
The satellite galaxies are typically distributed furthest from the center of the Galaxy, and are therefore arguably best placed to constrain its outer mass distribution, and total mass, without extrapolation.

We intentionally restrict our joint $M(<r)$ estimate from 20 to 80 kpc, reflecting the range of our data.
Some literature constraints are reported at greater distances, but the bulk of the halo data (and arguably the constraining power) lies within 100~kpc.
Among these distant measurements using halo stars, there is a hint of a ceiling, with results not breaking over $\sim$$8\times 10^{11}\Msun$.
However, the reported uncertainties associated with these results vary substantially, and systematic effects may have been neglected or understated.
The literature constraints using more distant satellite galaxies and globular clusters do not show this ceiling.
While we do not report an enclosed mass estimate beyond 80~kpc, for illustrative purposes we show an extrapolated NFW profile that adopts our joint \Mtwoc constraint and a concentration of $c_\text{200c} = 12$, a value consistent with the \empdf constraints reported in \citet{Zhou:2026empdf}, the NIMBLE-B constraints for both tracers, and the NIMBLE-A BHB constraint.
This extrapolated profile lies above the most distant constraints using halo stars, but is consistent with the higher estimates from satellites.

Lastly, in Figure~\ref{fig:m200-lit-comparison} we present a compilation of \Mtwoc measurements from the literature, as well as all results using DESI DR2 from this work.
As seen for the enclosed mass profile, our results are broadly consistent with literature estimates using a variety of halo tracers and methodologies (see the next Section for discussion on rotation curve estimates).
In terms of reported uncertainties, our joint DESI DR2 constraint is among the most competitive available, several of which \citep[e.g.,][]{Callingham:2019, Shen2022_h3mwmass, Sun:2023, Ibata:2024} are remarkably consistent with ours.
The strength of our estimate is in combining multiple well-tested and calibrated methodologies in a statistically robust way.

\subsection{Tension with the Galactic rotation curve} \label{sec:rotcurve-tension}

There is a growing disagreement between dynamical models of the Milky Way rotation curve and more distant tracers in the halo (see \citealp{Melchiorri:2026} for a pedagogical summary).
Beginning with \textit{Gaia} DR2 \citep{Eilers:2019} and especially with improved DR3 \citep{Jiao:2023, Wang:2023, Ou:2024}, the Milky Way's stellar disk shows a declining rotation curve, leading to claims of total masses that are substantially below nearly all estimates using halo tracers.

While our experimental setup is not designed to resolve this tension, our measurement is far more consistent with prior halo tracer estimates (Figures~\ref{fig:MencLit} and \ref{fig:m200-lit-comparison}).
All of our individual and combined measurements prefer an enclosed mass profile that continues to increase from $r = 20$~kpc to 80~kpc.
Furthermore, our joint $M(<r)$ profile also agrees with rotation curve measurements at $r = 20$~kpc \citep{Jiao:2023, Ou:2024}, diverging only at larger distances where the DESI halo stars continue.
Detailed analyses have shown that these rotation curve measurements can be biased due to survey selection functions, inaccurate asymmetric drift corrections, axis asymmetries, or disequilibrium in the Galactic disk \citep{Koop:2024, Ou:2025}.
In addition, such methods typically treat the circular velocity measured at different radii as independent measurements, resulting in mass model uncertainties that may be overconfident \citep{Oman:2024}.

Of the three proposed solutions to this tension outlined in Section 14 of \citet{Melchiorri:2026}, the DESI DR2 results strongly favor a scenario where the declining rotation curve is an artefact of these (or other) biases, and the true mass of the Milky Way is $\Mtwoc \gtrsim 1\times10^{12} \Msun$.
The ``compromise'' scenario, where the true mass of the Milky Way is $\Mtwoc = 5-8\times10^{11} \Msun$ (broadly consistent with stream-based estimates but below those using globular clusters or satellite galaxies) is nominally consistent with our joint $M(<r)$ profile\footnote{Ignoring our extrapolated (however well-calibrated) \Mtwoc estimates.}, but would require a rapid flattening of the enclosed mass profile beyond 80~kpc.
Based on our measurements, such flattening only seems possible from the NIMBLE-A RRL modeling (see Figure~\ref{fig:mw_mass_separate}).
However, we favor the interpretation that the NIMBLE-A RRL profile is adversely affected by the tracer density reconstruction (Section \ref{sec:dr2-results}).

Finally, we restrict our joint analysis of $M(<r)$ to beyond 20~kpc from the Galactic center.
Within this distance, assumptions about the inner potential dominate, including any baryonic contraction in the dark matter distribution \citep{Cautun2020, Dado:2026}.
In particular, we avoid any calculation of the local dark matter density using our data, which would require extrapolations to the solar neighborhood.
Constraints from direct detection experiments are also sensitive to assumptions in the velocity distribution of local dark matter, which may be affected by the Milky Way's unique assembly history, in particular the presence of a massive nearby LMC \citep{Besla:2019, Smith-Orlik:2023}.

\subsection{Are Milky Way mass estimates systematics-dominated?} \label{sec:systematics-regime}

When deriving our joint estimates, we find that the empirically determined systematic uncertainties for each method are generally larger than their reported internal dispersions.
This can be seen visually in Figure~\ref{fig:rel-menc-comparison}, where the scatter among measurements for a given method is typically larger than the reported uncertainties would suggest.
While the relative contribution to the total uncertainty varies both with distance and between methods, this suggests that individual modeling techniques (both here and likely in the literature) tend to be overconfident in their reported mass estimates.
Such systematic uncertainties could arise from deviations from common assumptions, including steady-state and spherical symmetry, and can not be reduced with the increase of observational sample size \citep{Wang:2017halomodeling,Wang+18,Han20tracer,Zhou:2026empdf}.
This highlights the need for large-scale efforts that consider multiple methods simultaneously and account for the variance inherent to galaxies that assemble hierarchically.

This finding raises a difficult question of whether future efforts should invest more into improving observational data or theoretical modeling techniques.
Within $r \lesssim 50 \kpc$, both the data samples and the empirically-determined biases are large enough that it is easier to argue for focusing on the theory.
Such improvements could include relaxing the assumptions of spherical symmetry and dynamical equilibrium, including moving beyond analytical distributions using flexible techniques like simulation-based inference.
Beyond $r \gtrsim 70 \kpc$, the smaller sample of halo stars with larger proper motion uncertainties becomes more limiting, particularly when estimating the total mass \Mtwoc.
This suggests that targeted spectroscopic follow-up of fainter BHBs and RRLs, expanding the set of tracers to include more luminous red giant stars and Milky Way satellites, and improved astrometric measurements expected from upcoming \textit{Gaia} data releases will all play a key role in next-generation measurements.

Finally, we note that neither our modeling methods nor our mock catalogues\footnote{While the Auriga simulations have massive satellites comparable to that of the LMC, they are located beyond $r = 100 \kpc$ at the present day and unlikely to perturb their stellar halo as strongly as the MW-LMC system.} explicitly capture dynamical perturbations due to the LMC.
A growing body of work has established that this massive ($\sim$10$^{11} \Msun$) satellite on a highly eccentric orbit \citep{Besla:2007, Kallivayalil:2013} induces large-scale distortions in the Milky Way's dark matter distribution \citep{Garavito-Camargo:2019}, shifting the barycenter of the Milky Way \citep{Gomez:2015} and producing a velocity dipole in the stellar halo \citep{Petersen:2021, Erkal:2021}.
Such effects can bias mass measurements based on equilibrium assumptions by up to 20~per~cent in the radial range spanned by our data \citep{Erkal:2020}, comparable to the biases we see in our mocks (e.g., Figure~\ref{fig:method-auridesi-summary}).
However, the \citet{Erkal:2020} analysis perturbs a stellar halo that is in a smooth and equilibrium state, ignoring the hierarchical nature of structure formation that results in stellar haloes rich with unrelaxed substructure \citep[e.g.,][]{Bullock:2005, Cooper:2010}.
While this work has focused on the dynamical effects of substructure generally, future efforts could address this using mocks based on cosmological simulations that capture LMC-like interactions \citep[e.g.,][]{Buch:2024}.

\subsection{Relationship to the individual measurements} \label{sec:relationship-individual-joint}

This work is accompanied by a set of supporting papers \citep{Zhou:2026empdf, Medina:2026gme,Valluri:2026nimble} that describe in detail the four individual methods, any calibration or validation performed using the mocks, and their results for the Milky Way.
Our joint estimate combines these mass measurements while accounting for biases and correlations between the methods.
This leverages the combined strengths of the individual methods, in pursuit of the best constraint available using the first three years of DESI data.

However, we caution that the joint estimate should not be read as a substitute for the individual measurements.
As it incorporates each individual estimate, it inherits (at least to some degree) the union of their modelling assumptions, which propagate into the final result.
While the joint estimate is calibrated against our suite of mocks, that calibration is only as reliable as the mocks themselves: it assumes that they faithfully reproduce the real Milky Way data, and that the systematic biases, uncertainties, and per-method weights are generic enough to transfer to the real observation.
To fully appreciate the strengths and limitations of these measurements, we advise consulting each supporting paper to weigh its assumptions and caveats individually, rather than treating the joint estimate as a single answer.

\section{Summary} \label{sec:summary}

We have presented Milky Way mass measurements using distant halo stars from the first three years of DESI operations that will be included in the upcoming DR2.
Our analysis includes four dynamical modeling methods: \empdf \citep{Zhou:2026empdf}, GME \citep{Medina:2026gme}, NIMBLE-A, and NIMBLE-B \citep{Valluri:2026nimble}.
Each of these techniques were validated using the same set of mock surveys, which were generated from cosmological simulations and incorporate selection functions and observational uncertainties that mirror the real DESI data \citep{Riley:2026auridesi}.
Based on the results from these mock tests, we combine the mass estimates using a robust statistical methodology that accounts for per-method biases, error floors, and correlations between methods.

Our main results are as follows:
\begin{itemize}
    \item Our joint estimate of the enclosed mass within 80~kpc and total virial mass of the Milky Way are $M(r < 80\kpc) = \JointMencEighty$ and $\Mtwoc = \JointMtwoc$, respectively.
    \item The joint mass estimates using only RRLs are in good agreement with those using only BHBs. For the individual methods post-correction, the two tracers broadly agree for \empdf and NIMBLE-B across all radii. GME and NIMBLE-A show the most difference between the two tracers, though the magnitude of the disagreement, and even which tracer yields a higher mass, varies with distance and method.
    \item In the absence of explicit calibration (as in \citealp{Zhou:2026empdf} for \empdf), none of our individual methods yield unbiased mass estimates across the full suite of mock surveys. In addition, the systematic uncertainties determined during our joint estimate procedure are larger than the uncertainties reported by each method. This highlights the need to quantify the impact of methodology assumptions using mocks that capture cosmological accretion histories and disequilibrium.
    \item Both our joint $M(<r)$ and \Mtwoc measurements strongly favor an enclosed mass profile that continues to rise beyond $r = 20$~kpc, agreeing with other studies using distant halo tracers but in tension with mass models of the Galactic rotation curve.
\end{itemize}

As discussed in Section~\ref{sec:relationship-individual-joint}, these results should not be read as replacing the individual mass estimates in the supporting papers \citep{Zhou:2026empdf, Medina:2026gme, Valluri:2026nimble}.
Importantly, our joint estimate relies on how well the lessons learned from our mocks (in particular, the biases and weights to assign to each method) transfer to the real Milky Way data.
Each supporting paper also discusses methodological improvements and tests on the mock catalogues, often including their own corrections to measurements based on those mock tests.

For external groups interested in modeling the data, the mocks are already publicly available (see \citealp{Riley:2026auridesi} for details) to develop and test methodologies, while the real Milky Way data will be released in the broader DESI DR2.

\begin{acknowledgements}
This material is based upon work supported by the U.S.~Department of Energy (DOE), Office of Science, Office of High-Energy Physics, under Contract No.~DE-AC02-05CH11231, and by the National Energy Research Scientific Computing Center, a DOE Office of Science User Facility under the same contract.
Additional support for DESI was provided by the U.S.~National Science Foundation (NSF), Division of Astronomical Sciences under Contract No.~AST-0950945 to the NSF's National Optical-Infrared Astronomy Research Laboratory; the Science and Technology Facilities Council of the United Kingdom; the Gordon and Betty Moore Foundation; the Heising-Simons Foundation; the French Alternative Energies and Atomic Energy Commission (CEA); the Secretariat of Science, Humanities, Technology and Innovation (SECIHTI) of Mexico; the Ministry of Science, Innovation and Universities of Spain (MICIU/AEI/10.13039/501100011033), and by the DESI Member Institutions: \url{https://www.desi.lbl.gov/collaborating-institutions}.
Any opinions, findings, and conclusions or recommendations expressed in this material are those of the author(s) and do not necessarily reflect the views of the U.~S.~National Science Foundation, the U.~S.~Department of Energy, or any of the listed funding agencies.

The authors are honored to be permitted to conduct scientific research on I'oligam Du'ag (Kitt Peak), a mountain with particular significance to the Tohono O'odham Nation.
\end{acknowledgements}

\section*{Data Availability} \label{sec:data-availability}

The data used in this analysis will be made public along with Data Release~2 (details in \url{https://data.desi.lbl.gov/doc/releases/}).
The data points corresponding to the figures from this paper are available from this Zenodo repository: \url{https://doi.org/10.5281/zenodo.23712179}.

\begin{contribution}
A.~Bystr\"{o}m and G.~E.~Medina assembled the BHB and RRL catalogues, respectively.
S.~E.~Koposov and G.~E.~Medina quantified the BHB and RRL selection functions, respectively.
A.~H.~Riley, N.~Kizhuprakkat, and A.~P.~Cooper produced the AuriDESI mock catalogues.
Y.~Zhou, J.~Han, W.~Wang, and Z.~Li conducted the \empdf analysis.
G.~E.~Medina, T.~S.~Li, and G.~M.~Eadie conducted the GME analysis.
M.~Valluri, T.~M.~Callingham, and N.~Rehemtulla conducted the NIMBLE analysis.
T.~M.~Callingham and O.~Y.~Gnedin designed and implemented the joint mass estimate.
L.~Beraldo e Silva, A.~Bystr\"{o}m, T.~M.~Callingham, A.~P.~Cooper, A.~Dey, C.~S.~Frenk, O.~Y.~Gnedin, J.~Han, S.~Li, G.~E.~Medina, A.~H.~Riley, M.~Valluri, W.~Wang, and Y.~Zhou drafted substantial original text.
T.~M.~Callingham, A.~H.~Riley, and M.~Valluri edited the text.
A.~H.~Riley and M.~Valluri were the coordinators of this Key Project.

The remaining authors have made contributions to this paper that include, but are not limited to, the construction of DESI and other aspects of collecting the data; data processing and calibration; developing broadly used methods, codes, and simulations; running the pipelines and validation tests; and promoting the science analysis.
All authors reviewed and approved the final manuscript.
\end{contribution}

\section*{Software}

This research made use of the Python programming language, along with many community-developed or maintained software packages including:
    \textsc{agama} \citep{agama},
    \textsc{arviz} \citep{arviz},
    \textsc{astropy} \citep{Astropy:2013, Astropy:2018, Astropy:2022},
    \textsc{cmasher} \citep{cmasher},
    \textsc{corner.py} \citep{corner},
    \textsc{cython} \citep{cython},
    \textsc{emcee} \citep{emcee},
    \textsc{gala} \citep{gala},
    \textsc{galpy} \citep{galpy},
    \textsc{h5py} \citep{h5py},
    \textsc{jupyter} \citep{ipython, jupyter},
    \textsc{matplotlib} \citep{matplotlib},
    \textsc{numpy} \citep{numpy},
    \textsc{numpyro} \citep{pyro, numpyro},
    \textsc{pandas} \citep{pandas},
    \textsc{scipy} \citep{scipy},
    \textsc{seaborn} \citep{seaborn},
    and \textsc{stan} \citep{Carpenter:2017, stan-manual}.
We also thank the maintainers of the \textsc{arepo-snap-util} package and the \textit{Gaia} Project Scientist Support Team and the \textit{Gaia} Data Processing and Analysis Consortium (DPAC).
Parts of the results in this work make use of the colormaps in the \textsc{cmasher} package.
Software citation information aggregated using \href{https://www.tomwagg.com/software-citation-station/}{The Software Citation Station} \citep{software-citation-station-paper}.

Use of large language models (LLMs) in this work was governed by internal DESI Collaboration policy that emphasizes disclosure and ultimate responsibility of the scientist.
Individual authors used LLM-based tools to assist in the following tasks: code generation (creating figures, centralizing figure data), code optimization (improving runtime), code debugging, and proofreading manuscript text.
LLMs used include Claude (Anthropic; Opus 5.5) and GPT (OpenAI; GPT-5, GPT-6).
All scientific conclusions are the authors' own.

\bibliographystyle{aasjournalv7.1}
\bibliography{main, dr2papers, software}

\begin{appendix}

\section{Results for all mocks} \label{app:all-mocks-figures}

\begin{figure*}
\centering
\includegraphics[width=\linewidth]{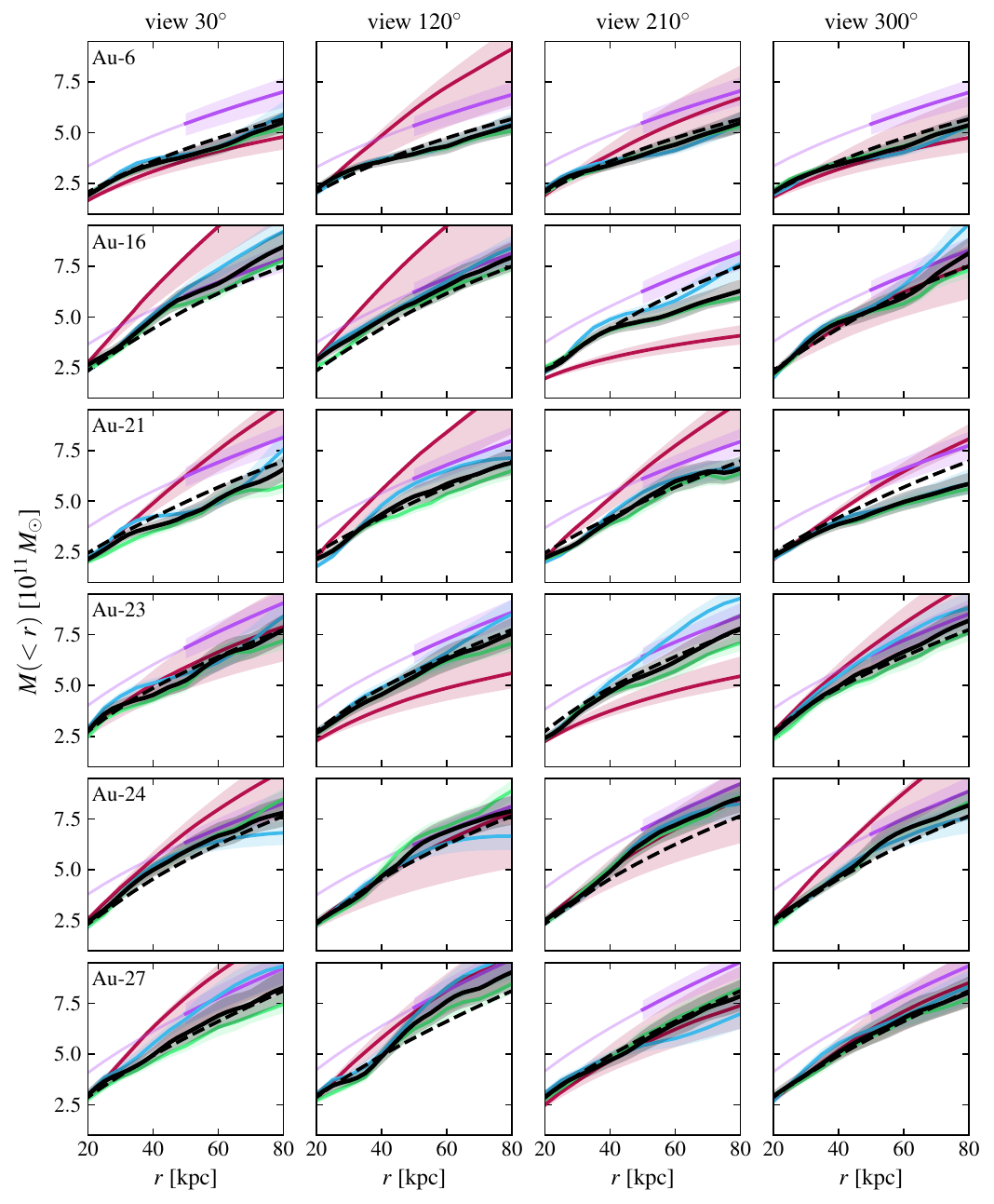}
\caption{Performance of the individual methods on all AuriDESI mock RRL catalogues, in a similar manner as the top panel of Figure~\ref{fig:menc-au6-comparison}.
}
\label{fig:menc-comparison-all-rrl}
\end{figure*}

\begin{figure*}
\centering
\includegraphics[width=\linewidth]{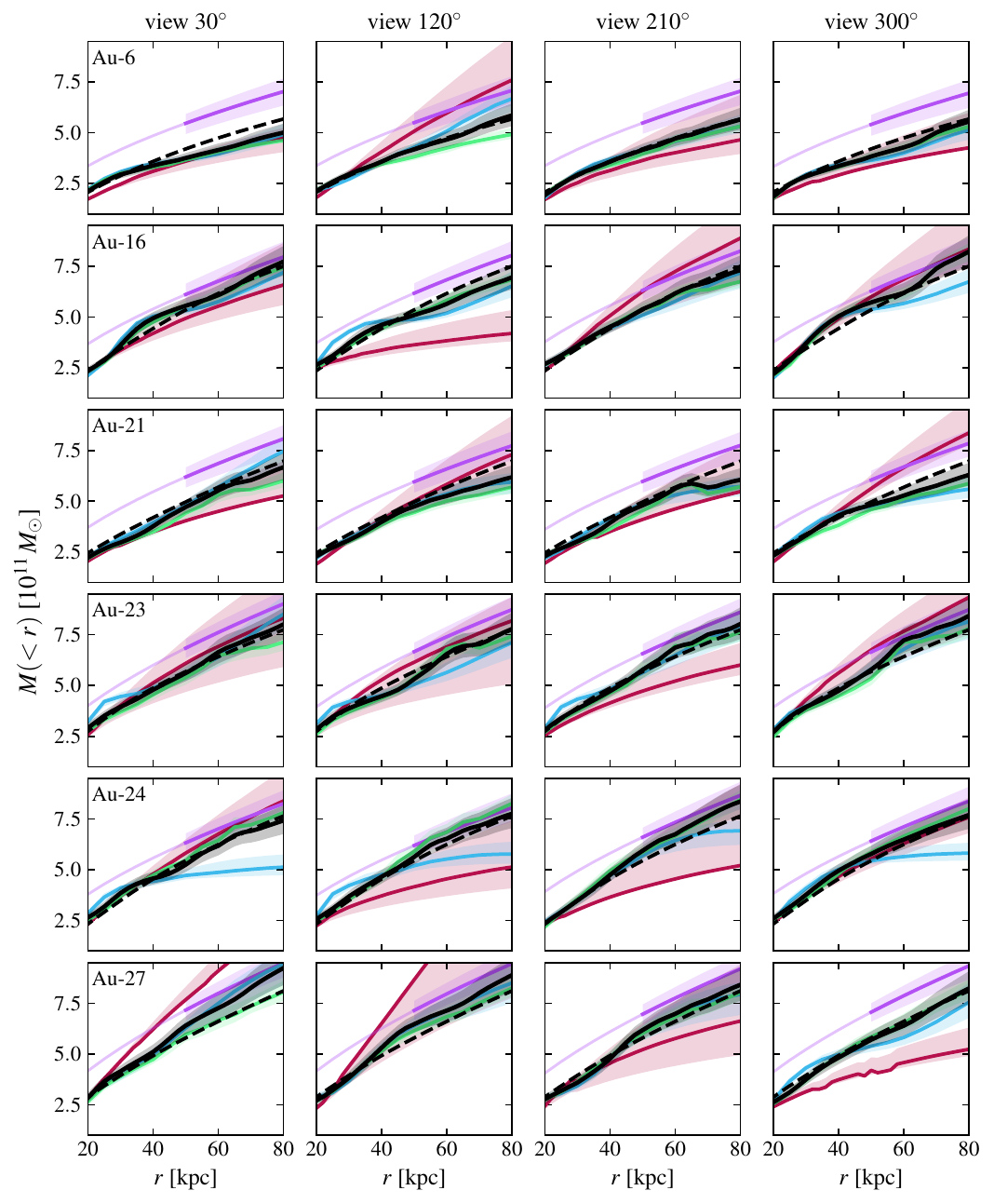}
\caption{Same as Figure~\ref{fig:menc-comparison-all-rrl}, but for all BHB mocks.
}
\label{fig:menc-comparison-all-bhb}
\end{figure*}

In Figure~\ref{fig:menc-comparison-all-rrl} we show the mass estimates for all mock RRL catalogues, in a manner similar to the top panel of Figure~\ref{fig:menc-au6-comparison}.
Figure~\ref{fig:menc-comparison-all-bhb} shows the same, but for all mock BHB catalogues.
As in Figure~\ref{fig:menc-au6-comparison}, the individual methods are not corrected for their respective biases, but the combined estimate does include these corrections.

\section{Combined mass estimate formalism} \label{app:combined-math-both}

In this Section we provide the mathematical formalism for the joint enclosed (Section \ref{app:mass-enclosed-fitting}) and total (Section \ref{App:M200Fitting}) mass estimates.
We have posterior distributions of $M(<r)$ and \Mtwoc{} from four methods (\empdf, GME, NIMBLE-A, NIMBLE-B) using two tracers each (RRL and BHB).
This gives us $N_m=8$ estimates at a given Galactocentric distance $r$ for $M(r)$, and \Mtwoc.
We calibrate the result of each method-tracer combination using the $N=24$ AuriDESI mocks (six halos observed at four viewing angles), treating each combination of the halo and viewing angle as an independent mock measurement.

We begin with the posterior PDF of mass provided by each method-tracer combination.
As described in the supporting papers, some methods evaluate the performance on mocks and then apply their own bias corrections.
We use the original PDF as directly calculated\footnote{With the exception of emPDF, where we use the estimate that includes a bias correction to account for measurement uncertainties.}, because here we evaluate the bias and include it in the joint estimate.
We find that majority of the mass PDF are better described by a log-normal distribution rather than a normal distribution, and therefore we work with the logarithmic mass variable: $z \equiv \log_{10}\!\left(M/10^{11}\,M_\odot\right)$.

At a given Galactocentric distance, we calculate the deviation of $\hat z_{m}$ from the true mass in each simulated galaxy and use this set to estimate the bias and systematic uncertainty of each method.
We define the residual of method $m$ on mock $i$ as
\begin{equation}
    r_{m,i} = \hat z_{m,i} - z_{{\rm true},i},
\end{equation}
where $\hat z_{m,i}$ is the raw measurement and $z_{{\rm true},i}$ is known true mass.

\subsection{Enclosed mass profile} \label{app:mass-enclosed-fitting}

The PDF are typically asymmetric around the median, either in linear or log mass. To capture the shape of the PDF more accurately, we parameterize them as a split-Gaussian distribution:
\begin{equation}
  p(z | \hat z_m,\sigma_m) = \frac{\sqrt{2/\pi}}{\sigma_{m,-} + \sigma_{m,+}}
  \exp\!\left[-\frac{(z-\hat z_m)^2}{2\,\sigma_m(z)^2}\right],
\end{equation}
where $\hat z_m$ is the peak of each method's PDF, and the asymmetric standard deviation takes two values:
\begin{equation}
  \sigma_m(z)=
  \begin{cases}
  \sigma_{m,-} & \text{if } z < \hat z_m,\\
  \sigma_{m,+} & \text{if } z \ge \hat z_m.
  \end{cases}
\end{equation}

We determine $\sigma_{m,-}$ and $\sigma_{m,+}$ by fitting a split-Gaussian distribution to percentile values of each PDF.

The bias of method $m$ is estimated as
\begin{equation}
    \hat\Delta_m = \frac{1}{N} \sum_{i=1}^{N} r_{m,i}
\end{equation}
Then we use the residual variance
\begin{equation}
    S_{mn} = \frac{1}{N-1}\sum_{i=1}^{N}\, (r_{m,i}-\hat\Delta_m)\, (r_{n,i}-\hat\Delta_m)
\end{equation}
to estimate the covariance between the methods. Because we have an insufficient number of mocks ($N=24$) to estimate $N_m (N_m+1)/2 = 36$ elements of the covariance matrix, we apply a technique of shrinkage toward the diagonal:
\begin{equation}
    \mathbf{S}_{\rm res}
    = (1-\lambda)\,\mathbf{S} + \lambda\,\mathrm{diag}(\mathbf{S}),
\end{equation}
where the parameter $\lambda$ controls the sensitivity to noisy off-diagonal covariance estimates.
Here \(\lambda=0\) keeps the full covariance, whereas \(\lambda=1\) discards all off-diagonal correlations.
We take $\lambda = 0.5$, based on a leave-one-out validation on the mocks, as described in Section~\ref{sec:mass_enclosed_method}.

The correlation matrix estimated from the residuals is
\begin{equation}
    R_{mn} = \frac{(S_{\rm res})_{mn}}
    {\sqrt{(S_{\rm res})_{mm}\,(S_{\rm res})_{nn}}}.
\end{equation}
This encodes both statistical correlations (arising from shared tracer data) and systematic correlations (arising from shared sensitivity to, e.g., halo structure or non-equilibrium).
The two correlations are comparable in strength.
They must be included in the joint estimate, because they increase the uncertainty of the derived mass.

We also notice that different methods carry significantly different biases in the mocks, which indicate different sensitivity to their adopted assumptions.
These bias values may still underestimate the true bias in the Milky Way measurements, because of the presence of large velocity outliers (above 500 km/s) in the DESI data but not in the mock samples.
To reduce sensitivity of the joint estimate to a single statistically precise but possibly inaccurate estimate, we include a fixed method-dependent floor $f_m$ (same for both tracers) on the systematic uncertainty.
It enters as the diagonal matrix $F_{mm} = f_m^2$.
This floor allows us to balance the effective contributions of the individual methods to the combined estimate.
The full systematic covariance is
\begin{equation}
 \mathbf{C}_{\rm sys} = \mathbf{S}_{\rm res} + \mathbf{F}.
\end{equation}

Finally, since the bias vector \(\hat{\boldsymbol\Delta}\) is estimated from a finite number of mocks, our bias calibration carries the associated uncertainty \(\mathbf C_{\rm bias} \approx \frac{1}{N} \mathbf S_{\rm res}\).
In principle this covariance term should be propagated into the final Milky Way mass estimate.
However, we find that this term is completely subdominant to the other contributions to \(\mathbf{C}_{\rm sys}\), and do not include it for simplicity.

The joint constraint on the Milky Way mass is drawn from a multivariate Gaussian distribution.
Given a set of bias-corrected estimates for the DESI data
\begin{equation}
    y_m = \hat z_m - \hat\Delta_m,
\end{equation}
the likelihood for the true log mass $z$ is
\begin{equation}
    -2\ln\mathcal{L}(z)
    = \boldsymbol\delta(z)^{\!\top}\,
      \mathbf{C}_{\rm tot}^{-1}(z)\,
      \boldsymbol\delta(z)
    + \ln\bigl|\mathbf{C}_{\rm tot}(z)\bigr|,
\end{equation}
where
\begin{equation}
    \delta_m(z) = y_m - z,
\end{equation}
and the total covariance is
\begin{equation}
    \mathbf{C}_{\rm tot}(z)
    = \mathbf{C}_{\rm stat}(z) + \mathbf{C}_{\rm sys}.
\end{equation}
Here \(\mathbf C_{\rm stat}\) is the statistical covariance of the Milky Way measurements, and \(\mathbf C_{\rm sys}\) is the mock-calibrated residual covariance, including correlations between methods.
The statistical covariance retains the split-Gaussian structure:
\begin{equation}
    C_{{\rm stat},mn}(z)
    = \tilde\sigma_m(z)\;\tilde\sigma_n(z)\;R_{mn},
\end{equation}
with
\begin{equation}
    \tilde\sigma_m(z) =
    \begin{cases}
        \sigma^-_m & \text{if } z < y_m, \\
        \sigma^+_m & \text{if } z \ge y_m.
    \end{cases}
\end{equation}

We obtain the maximum-likelihood estimate $\hat z$ by bounded numerical minimization of $-2\ln\mathcal{L}(z)$.
The 68~per~cent confidence interval $[z_{\rm lo},\, z_{\rm hi}]$ is defined by
\begin{equation}
    -2\ln\mathcal{L}(z_{\rm lo,hi})
    = -2\ln\mathcal{L}(\hat z) + 1,
\end{equation}
and solved by numerical root-finding on each side of the maximum.

\subsection{Total mass} \label{App:M200Fitting}

Each method's result gives a chain of $M_{200}$ values that we summarise with a median and standard deviation from percentiles.
For method $m$ on mock $i$ we have
\begin{align}
    \hat z_{m,i} &= \mathrm{median}_k\,\bigl(\log_{10} M_{200}^{(k)}\bigr),
    \label{eq:z_summary}\\
    \sigma_{m,i} &= \tfrac12\bigl[Q_{84}(\log_{10} M_{200}) - Q_{16}(\log_{10} M_{200})\bigr].
    \label{eq:sigma_summary}
\end{align}
 
Each estimator's mock $i$ observation  is modelled as
\begin{equation}
    \hat z_{m,i} \;=\; z_{\mathrm{true},i} \;+\; \Delta_m \;+\; \varepsilon_{m,i},
    \label{eq:forward_model_mock}
\end{equation}
where $\Delta_m$ is a method-specific bias, and the error vector
$\boldsymbol\varepsilon_i = (\varepsilon_{1,i},\ldots,\varepsilon_{M,i})^{\!\top}$
is drawn from a multivariate Gaussian,
\begin{equation}
    \boldsymbol\varepsilon_i \;\sim\; \mathcal{N}_M\!\bigl(\mathbf{0},\;\boldsymbol\Sigma_i\bigr),
    \label{eq:error_model}
\end{equation}
with covariance
\begin{equation}
    \boldsymbol\Sigma_i \;=\;
        \mathrm{diag}\!\bigl(\boldsymbol\sigma_{\mathrm{eff},i}\bigr)\;
        \mathbf{R}\;
        \mathrm{diag}\!\bigl(\boldsymbol\sigma_{\mathrm{eff},i}\bigr).
    \label{eq:Sigma_mock}
\end{equation}
The per-method, per-mock effective scatter is
\begin{equation}
    \sigma_{\mathrm{eff},m,i}^{2} \;=\;
        f_m^{2} \;+\; \bigl(s_m\,\sigma_{m,i}\bigr)^{2},
    \label{eq:sigma_eff}
\end{equation}
where $f_m$ method floor uncertainty that absorbs systematics (deviations from the assumptions of spherical symmetry, equilibrium, or adopted mass profile).
$s_m$ is a scaling of the quoted statistical error $\sigma_{m,i}$.
$\mathbf{R}$ is a single $M\!\times\!M$ correlation matrix shared across all mocks,
describing both shared statistical correlation (e.g., the results using RRLs or BHBs) and shared systematic correlation (methods sensitive to the same physical effects).
We restrict this to be strictly positive correlations, as there are no physical reasons for anti-correlations between our estimators; disequilibrium or other systematic effects will bias correlated methods in the same direction.

The residual vector for mock $i$ is then distributed normally:
\begin{align}
    \hat{\mathbf{r}}_i &\;=\;
        \hat{\mathbf{z}}_i
        - z_{\mathrm{true},i}\,\mathbf{1}
        - \boldsymbol\Delta
    \;\sim\;
        \mathcal{N}_M\!\bigl(\mathbf{0},\;\boldsymbol\Sigma_i\bigr).
    \label{eq:residual_likelihood}
\end{align}
 
This model is fit using the NumPyro \citep{pyro, numpyro} NUTS sampler with the following broad priors:
\begin{itemize}
    \item $\Delta_m \sim \mathcal{N}(0,\,0.3)$ --- biases up to a factor
          $\sim$2 in linear $M_{200}$.
    \item $f_m \sim \mathcal{N}(0,0.2) \geq 0$ (half-normal)
    \item $\log s_m \sim \mathcal{N}(0,\,0.5)$ 
    \item $\mathbf{R} \sim \mathrm{LKJ}(\eta=1)$ --- uniform on correlation
          matrices, with a steep quadratic penalty enforcing $R_{m,n} \geq 0$.
\end{itemize}

Once fit, each method's estimate on a mock or the Milky Way can be bias-corrected, with calibrated uncertainty $\sigma_{\mathrm{eff}}$:
\begin{align}
    \hat z_{\mathrm{cal},m,i} &\;=\; \hat z_{m,i} - \Delta_m,
    \label{eq:z_cal_mock}\\[4pt]
    \sigma_{\mathrm{cal},m,i}^{2} &\;=\; f_m^{2} + (s_m\,\sigma_{m,i})^{2}.
    \label{eq:sigma_cal_mock}
\end{align}
 
To find a joint estimate, we  combine the individual estimates by using weights $\mathbf{w}$:
\begin{equation}
    \hat z_{\mathrm{comb}}
        \;=\; \sum_m w_m\,\hat z_{\mathrm{cal},m},
    \label{eq:zcomb}\\[4pt]
\end{equation}

The weights are chosen to minimise the final variance:
\begin{align}
    \mathbf{w} &\;=\; \arg\min_{\mathbf{w}}\;
        \mathbf{w}^{\!\top}\boldsymbol\Sigma_{\mathrm{reg}}\,\mathbf{w}
    \quad
    \text{with}\quad
    \sum_m w_m = 1,\;\; w_m \geq 0.
    \label{eq:weights_qp}
\end{align}

To guard against overconfidence in the mock-calibrated relative precisions, the weights are computed from a regularised covariance that adds an independent additional systematic $\sigma_{\mathrm{extra}}$ to the diagonal:
\begin{equation}
    \boldsymbol\Sigma_{\mathrm{reg}} \;=\;
        \boldsymbol\Sigma
        \;+\; \sigma_{\mathrm{extra}}^{2}\,\mathbf{I},
    \label{eq:Sigma_reg}
\end{equation}
with $\sigma_{\mathrm{extra}} = 0.08$\,dex ($\sim$20~per~cent in linear $M_{200}$).
As $\sigma_{\mathrm{extra}}$ only enters the diagonal, it partially decorrelates methods for weight purposes; two highly correlated methods with similar $\sigma_{\mathrm{eff}}$ receive near-equal weights rather than one estimate dominating from overfitting.

In practice, we take the parameters from the chains $(\boldsymbol\Delta^{(s)}, \mathbf{f}^{(s)}, \mathbf{s}^{(s)}, \mathbf{R}^{(s)})$, then compute the weights and $z_{\mathrm{comb}}^{(s)}$ for each sample.
The chain $\{z_{\mathrm{comb}}^{(s)}\}$ is the full marginalised posterior; the reported uncertainty is taken from its 16th and 84th percentiles:
\begin{equation}
    M_{200} = P_{50}
        \;{}^{+\,(P_{84} - P_{50})}_{-\,(P_{50} - P_{16})}.
    \label{eq:mw_result}
\end{equation}
All uncertainties (bias correction, correlations, and calibration) are included in the chain.

In the Milky Way the methods all give raw estimates $(\hat z_{m}^{\mathrm{MW}}, \sigma_{m}^{\mathrm{MW}})$, which we apply our mock-based calibration to.
The Milky Way covariance uses the same correlation structure $\mathbf{R}$ as the mock calibration, with the effective scatter now including an additional bias correction uncertainty:
\begin{equation}
    \boldsymbol\Sigma^{\mathrm{MW}} \;=\;
        \mathrm{diag}\!\bigl(\boldsymbol\sigma_{\mathrm{eff}}^{\mathrm{MW}}\bigr)\,
        \mathbf{R}\,
        \mathrm{diag}\!\bigl(\boldsymbol\sigma_{\mathrm{eff}}^{\mathrm{MW}}\bigr),
    \label{eq:Sigma_mw}
\end{equation}
where
\begin{equation}
    \sigma_{\mathrm{eff},m}^{\mathrm{MW},2} \;=\;
        f_m^{2}
        \;+\; \bigl(s_m\,\sigma_{m}^{\mathrm{MW}}\bigr)^{2}
        \;+\; \Delta_m^{2}.
    \label{eq:sigma_eff_mw}
\end{equation}
The third term is the bias correction uncertainty; we apply the correction but give relative uncertainty to whether it transfers from AuriDESI to the Milky Way, since our methods are calibrated on a single mock suite.
Methods correlated in $\mathbf{R}$ (e.g.\ the same method on two tracers) will have correlated bias correction failures.
If the calibration is wrong for one, it is likely wrong for the other in the same direction.

\section{References for literature compilation} \label{app:lit-compilation-references}

The references for the literature Milky Way mass measurements in Figures~\ref{fig:MencLit} and \ref{fig:m200-lit-comparison} are shown in Table~\ref{tab:lit-references}.

\begin{deluxetable}{llll}
\tablecaption{Literature references corresponding to Figures~\ref{fig:MencLit} and \ref{fig:m200-lit-comparison}.}
\label{tab:lit-references}
\tablehead{
\colhead{Label} &
\colhead{Reference} &
\colhead{Label} &
\colhead{Reference}
}
\startdata
Ablimit17 & \citet{Ablimit:2017} &
Battaglia05 & \citet{Battaglia:2005} \\
Bhattacharjee14 & \citet{Bhattacharjee:2014} &
Bird22 & \citet{Bird:2022} \\
Callingham19 & \citet{Callingham:2019} &
Cautun20 & \citet{Cautun2020} \\
CorreaMagnus22 & \citet{CorreaMagnus:2022} &
Deason12a & \citet{Deason:2012broken} \\
Deason12b & \citet{Deason:2012coldveil} &
Deason19 & \citet{Deason:2019} \\
Deason21 & \citet{Deason2021} &
Eadie19 & \citet{Eadie:2019} \\
Erkal19 & \citet{Erkal:2019} &
Fritz20 & \citet{Fritz:2020} \\
Gnedin10 & \citet{Gnedin:2010} &
Grand19 & \citet{Grand:2019mwmass_tail} \\
Hattori21 & \citet{Hattori2021} &
Ibata24 & \citet{Ibata:2024} \\
Jiao23 & \citet{Jiao:2023} &
Kafle12 & \citet{Kafle:2012} \\
Karukes20 & \citet{Karukes:2020} &
Koposov23 & \citet{Koposov2023} \\
Kravtsov24 & \citet{Kravtsov:2024} &
Li20 & \citet{Li2020} \\
Li25 & \citet{LiZhaozhou2025} &
Makarov25 & \citet{Makarov:2025} \\
Malhan19 & \citet{Malhan:2019} &
McMillan17 & \citet{McMillan:2017} \\
Monari18 & \citet{Monari:2018} &
Nesti13 & \citet{Nesti:2013} \\
Ou24 & \citet{Ou:2024} &
Patel17 & \citet{Patel:2017} \\
Patel18 & \citet{Patel:2018} &
Posti19 & \citet{Posti:2019_mwmass} \\
Prudil22 & \citet{Prudil2022} &
Reino21 & \citet{Reino:2021} \\
Roche24 & \citet{Roche:2024} &
Shen22 & \citet{Shen2022_h3mwmass} \\
Slizewski22 & \citet{Slizewski:2022} &
SylosLabini23 & \citet{SylosLabini:2023} \\
SylosLabini24 & \citet{SylosLabini:2024} &
Sohn18 & \citet{Sohn:2018} \\
Sun23 & \citet{Sun:2023} &
Vasiliev19 & \citet{Vasiliev:2019_gcdynamics} \\
Vasiliev21 & \citet{Vasiliev:2021} &
Watkins19 & \citet{Watkins:2019} \\
Wegg19 & \citet{Wegg_2019} &
Williams15 & \citet{Williams:2015} \\
Williams17 & \citet{Williams:2017} &
Wu25 & \citet{Wu:2025} \\
Xue08 & \citet{Xue:2008} &
Zaritsky20 & \citet{Zaritsky:2020} \\
Zhai18 & \citet{Zhai:2018} &
Zhai20 & \citet{Zhai:2020} \\
Zhang25 & \citet{Zhang:2025} &
Zhou23 & \citet{Zhou:2023} \\
\enddata
\end{deluxetable}

\end{appendix}

\end{document}